\documentclass[aps,pra,reprint,amsmath,amssymb,
   floatfix,showpacs,showkeys,preprintnumbers,
   ]{revtex4-1}

\usepackage{hyperref}
\usepackage{graphicx}
\usepackage{dcolumn}
\usepackage{bm}

\newcommand{\DeltaAver}{\ensuremath{\overline{\Delta}}}
\newcommand{\DeltaMin}{\ensuremath{\Delta_{\mathrm{min}}}}
\newcommand{\chidosred}{\ensuremath{\chi^2_{\mathrm{red}}}}
\newcommand{\Bresol}{\ensuremath{\delta B_{\mathrm{res}}}}

\begin{document}

\title{Spectral statistics of molecular resonances in erbium isotopes: How chaotic are they?}

\author{Jordi Mur-Petit}
\email{jordi.mur@csic.es}
\altaffiliation{Present address: Clarendon Laboratory, University of Oxford, Parks Road, Oxford OX1 3PU, United Kingdom}
\affiliation{Instituto de Estructura de la Materia, IEM-CSIC, Serrano 123, 28006
  Madrid, Spain}

\author{Rafael A.\ Molina}
\email{rafael.molina@csic.es}
\affiliation{Instituto de Estructura de la Materia, IEM-CSIC, Serrano 123, 28006
  Madrid, Spain}

\begin{abstract}
We perform a comprehensive analysis of the spectral statistics of the molecular resonances in $^{166}$Er and $^{168}$Er observed in recent ultracold collision experiments [Frisch et al., Nature {\bf 507}, 475 (2014)] with the aim of determining the chaoticity of this system.
We calculate different independent statistical properties to check
their degree of agreement with random matrix theory (RMT), and analyze if they are consistent with the possibility of having missing resonances.
The analysis of the short-range fluctuations as a function of the magnetic field points to a steady increase of chaoticity until $B \sim 30$~G.
The repulsion parameter decreases for higher magnetic fields, an effect that can be interpreted as due to missing resonances. 
The analysis of long-range fluctuations allows us to be more quantitative and estimate a $20-25\%$ fraction of missing levels.
Finally, a study of the distribution of resonance widths provides additional evidence supporting missing resonances of small width compared with the experimental magnetic field resolution.
We conclude that further measurements with increased resolution will be necessary to give a final answer to the problem of missing resonances and the agreement with RMT.
\end{abstract}



\maketitle

%%%%%%%%%%%%%%%%%%%%%%%%%%%%%%%%%%%%%%%%%%%%%%%%%%%%%%%%
%%%%%%%%%%%%%%%%%%%%%%%%%%%%%%%%%%%%%%%%%%%%%%%%%%%%%%%%
%%%%%%%%%%%%%%%%%%%%%%%%%%%%%%%%%%%%%%%%%%%%%%%%%%%%%%%%

\section{Introduction}\label{sec:intro}

Understanding the structure of a molecule is a complex task that usually requires input from both experimental spectral data and electronic structure calculations. This joint effort has the goal of identifying the relevant symmetries and quantum numbers, ultimately allowing one to translate an initial apparently random collection of spectral lines into a coherent set of vibrational bands, rotational lines, etc~\cite{Bernath,BrownCarr}, and is analog to the process necessary to understand atomic or nuclear spectra.

In atomic and, especially, nuclear spectroscopy, it has long been recognized that some systems feature very complex spectra which defy this procedure. In essence, interactions between the different particles in the system are so strong that the many-body system becomes strongly correlated. The analysis of its spectrum can not identify each line unambiguously with a dominant set of quantum numbers, and one resorts instead to statistical studies.
Such chaotic quantum systems present very specific statistical properties in their spectra and wave functions that can be traced to results from random matrix theory (RMT) \cite{Haake_book,Stoeckmann_book}.
A limiting case of particular relevance holds when the quantum system has a fully chaotic classical analog. In this case, according to the original Bohigas-Giannoni-Schmit (BGS) conjecture, its spectrum will show ``the same fluctuation properties as predicted by the Gaussian orthogonal ensemble (GOE)'' \cite{Bohigas84}. This conjecture has now been given firm grounds at the semiclassical level \cite{Mueller04}.

There is an enormous amount of numerical and experimental evidence in support of the relationship between the properties of chaotic
systems and the properties of random matrices, both in systems with classical analog and in many-body quantum systems \cite{Stoeckmann_book,Gomez11}.
The most stringent tests so far have been performed on large data sets of nuclear levels~\cite{Weiden09,Gomez11} as well as on microwave billiards~\cite{Stoeckmann_book}. There exist also several studies on the quantum-chaotic character of small molecules spectra from stimulated-emission pumping spectroscopy~\cite{Pique1987,Zimmermann1988}. In the last decade, atomic and molecular physics has been enriched by new experimental and theoretical tools for high-resolution spectroscopy at very low energies~\cite{weiner1999,stuhl2014} as well as novel control and measurement protocols~\cite{krems2008,quemener2012,murpetit2012,murpetit2015}.
In this context, the field of quantum chaos has recently received new input from ultracold collision experiments~\cite{Frisch14}. In this work we present a detailed study of these new data.

Analyzing experimental systems to check whether they behave quantum chaotically or not is far from a straightforward process as several effects may lead to deviations from RMT predictions. 
First, many systems present intermediate behavior between regularity and chaos.
Second, the statistical analysis to compare with RMT results should be done for complete sequences of levels with the same quantum numbers.
There have been a number of papers on how to analyze incomplete or imperfect spectra \cite{Liou72,Zimmermann1988,Bohigas06,Molina07}. They study both short and long range spectral fluctuations. Here, by incomplete spectra we mean that there is a fraction of missing levels unaccounted for experimentally, and by imperfect spectra we mean that there are wrong assignments of quantum numbers in the experimental spectra. Crucially, the effects of both missing levels and imperfect spectra are similar on short-range fluctuations, reducing the level repulsion and making them appear non-chaotic. On the other hand, their effects are very different for long range fluctuations, which can be then used to discern these two effects. 
For example, Ref. \cite{Molina07} showed how the power spectrum is a useful statistic to discriminate between them, noting that missing levels affect low frequencies (i.e., correlations with longer range) first, while mixed symmetries affect high frequencies (correlations with shorter range) first.

Very general formulas can be derived for the power spectrum as a function of the number of mixed symmetries, the number of levels in each symmetry, and the fraction of missing levels for each symmetry. In practice, in order to have simple formulas useful to extract the fraction of missing levels or the number of mixed symmetries from the experimental data, assumptions about the original spectra are needed and the mentioned references assume that the original spectra follow the predictions of RMT, i.e., that the underlying system is fully chaotic. What would happen if this were not the case and the original spectra had intermediate statistics? Likely, one will obtain a result closer to Poisson statistics (i.e., regular dynamics) than what the chaoticity of the original spectra would suggest and that will be very difficult to interpret. However, the fact that short- and long-range correlations give mainly independent information can be used to our advantage.

Arguments for the study of different statistical properties comparing short- and long-range level statistics with wave function statistics have been given before, for example noting that the transition between regular and chaotic spectra is not universal and can reflect in different ways in the different statistics~\cite{Molina00}.
Incomplete and imperfect spectra are other instances when complementary approaches are needed.
For example, wave-function statistics give another completely independent test of the data versus RMT predictions and, if the missing levels are the ones with smaller widths, as it is expected for finite experimental resolution, one can obtain this fraction from the analysis of the statistical distribution of level \textit{widths} in comparison with the Porter-Thomas prediction of RMT \cite{Froehner80}.

The general strategy to follow is, then, to study as many independent statistical properties as possible and not restrict oneself to only short-range (nor only long-range) correlations. If the complete analysis gives a consistent value for the number of missing levels, one can conclude that this number reflects the experimental results and that the original data do follow RMT. Otherwise, several effects may be playing a relevant role, and the original data probably follow intermediate statistics.

Here, we apply such a comprehensive approach to the data on molecular resonances in a lanthanide atom recently reported by Frisch {\em et al.}~\cite{Frisch14}.
The authors of that work studied collisions of trapped cold erbium atoms as a function of the magnetic field \cite{Frisch14}. The Feshbach resonances of bosonic isotopes $^{166}$Er and $^{168}$Er were studied with high-resolution trap-loss spectroscopy. A statistical analysis was made of the positions of the resonances using the tools of RMT. The behavior obtained was intermediate between Poisson and Wigner-Dyson (WD) statistics although somewhat closer to WD. From these results it was concluded that the dynamics of collisions and formation of Er$_2$ molecules is very complex and presents characteristics of quantum chaos.
We have extended the analysis of the experimental data in Ref.~\cite{Frisch14}, performing a study of short and long-range fluctuations in level positions as well as of level widths.

The paper is organized as follows.
In Sec.~\ref{sec:levels} we present a detailed analysis of the short and long range fluctuations of level statistics. Our conclusions here agree generally with those in Ref.~\cite{Frisch14}. At the same time, our results enable an analysis of the magnetic field dependence of the degree of chaoticity, leading us to a quantitative account of the origin of the deviations from Wigner-Dyson of both the short- and long-range correlations in the spectra. In addition, our calculations allow to estimate a fraction of missing levels at high magnetic fields.
Then, in Sec.~\ref{sec:widths} we present calculations of the level width statistics. From this study, it follows that there is a substantial fraction of missing levels with small widths\textemdash i.e., with widths below the current magnetic field resolution\textemdash which precludes reaching definitive conclusions on the chaoticity of the system.
The conclusions from these studies are summarized in Sec.~\ref{sec:concl}.

%%%%%%%%%%%%%%%%%%%%%%%%%%%%%%%%%%%%%%%%%%%%%%%%%%%%%%%%
%%%%%%%%%%%%%%%%%%%%%%%%%%%%%%%%%%%%%%%%%%%%%%%%%%%%%%%%
%%%%%%%%%%%%%%%%%%%%%%%%%%%%%%%%%%%%%%%%%%%%%%%%%%%%%%%%
\section{Statistics of energy levels}\label{sec:levels}

%%%%%%%%%%%%%%%%%%%%%%%%%%%%%%%%%%%%%%%%%%%%%%%%%%%%%%%%
%%%%%%%%%%%%%%%%%%%%%%%%%%%%%%%%%%%%%%%%%%%%%%%%%%%%%%%%
\subsection{Unfolding the spectra}\label{ssec:unfolding}

The unfolding of the spectra is the procedure by which the eigenenergies of the system $E_i$ are transformed to new unfolded energies $\epsilon_i$ so the density of levels is locally constant and equal to unity \cite{Haake_book}. It is a heuristic procedure, essential in order to compare the statistical analysis of the spectra with RMT results. Doing the unfolding erroneously can give rise to misleading signatures of chaos \cite{Gomez02}.
Fortunately, for the erbium isotopes it is relatively uncomplicated. 
In Ref. \cite{Frisch14} it was noticed that the density of resonances was constant for magnetic fields $B\gtrsim 30$~G. The unfolding in that case is trivial and the results for the spacing distribution and other statistical measures very reliable.

\begin{figure}[tb]
 \centering
 \includegraphics[width=\linewidth]{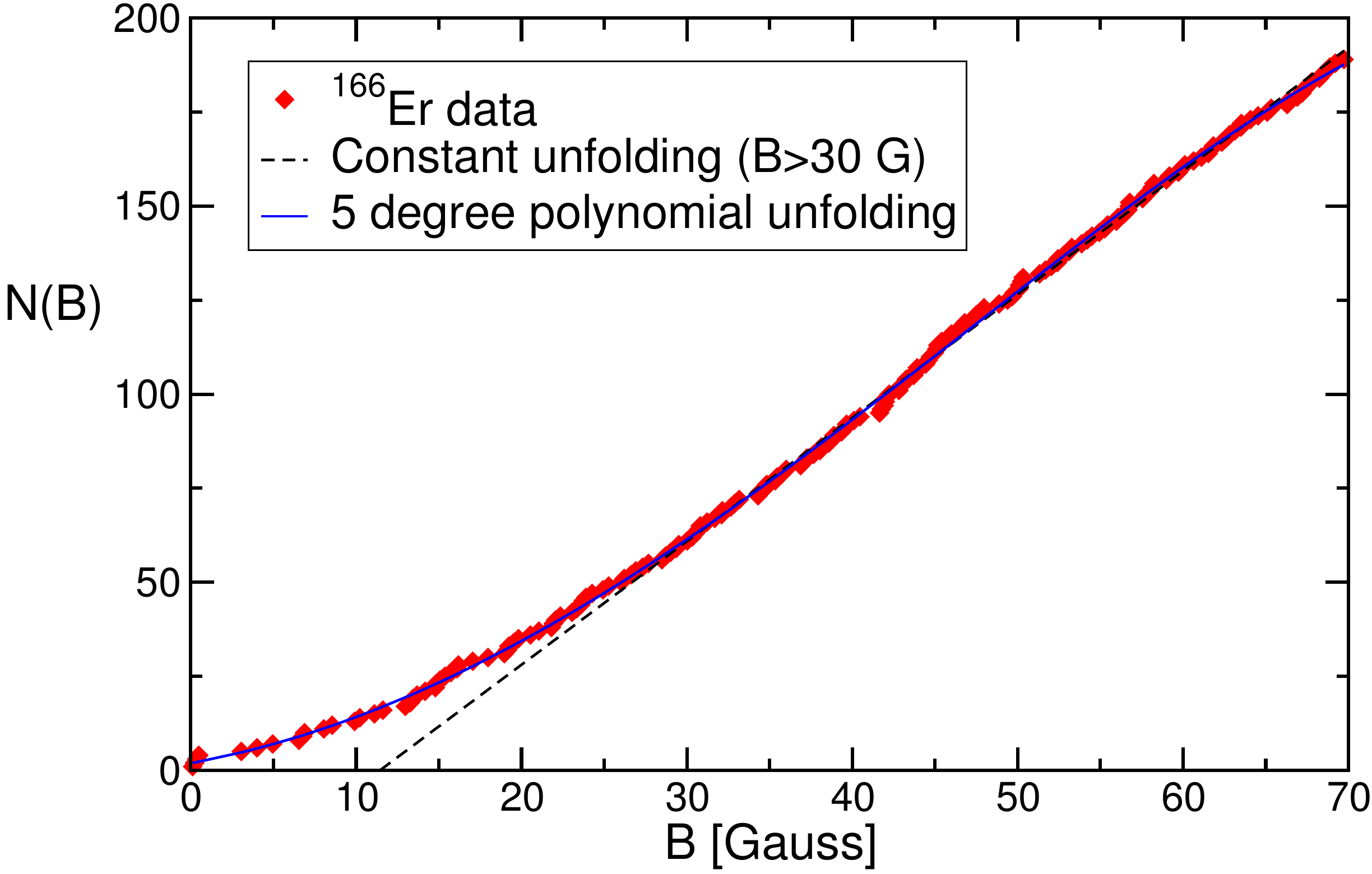}
 \includegraphics[width=\linewidth]{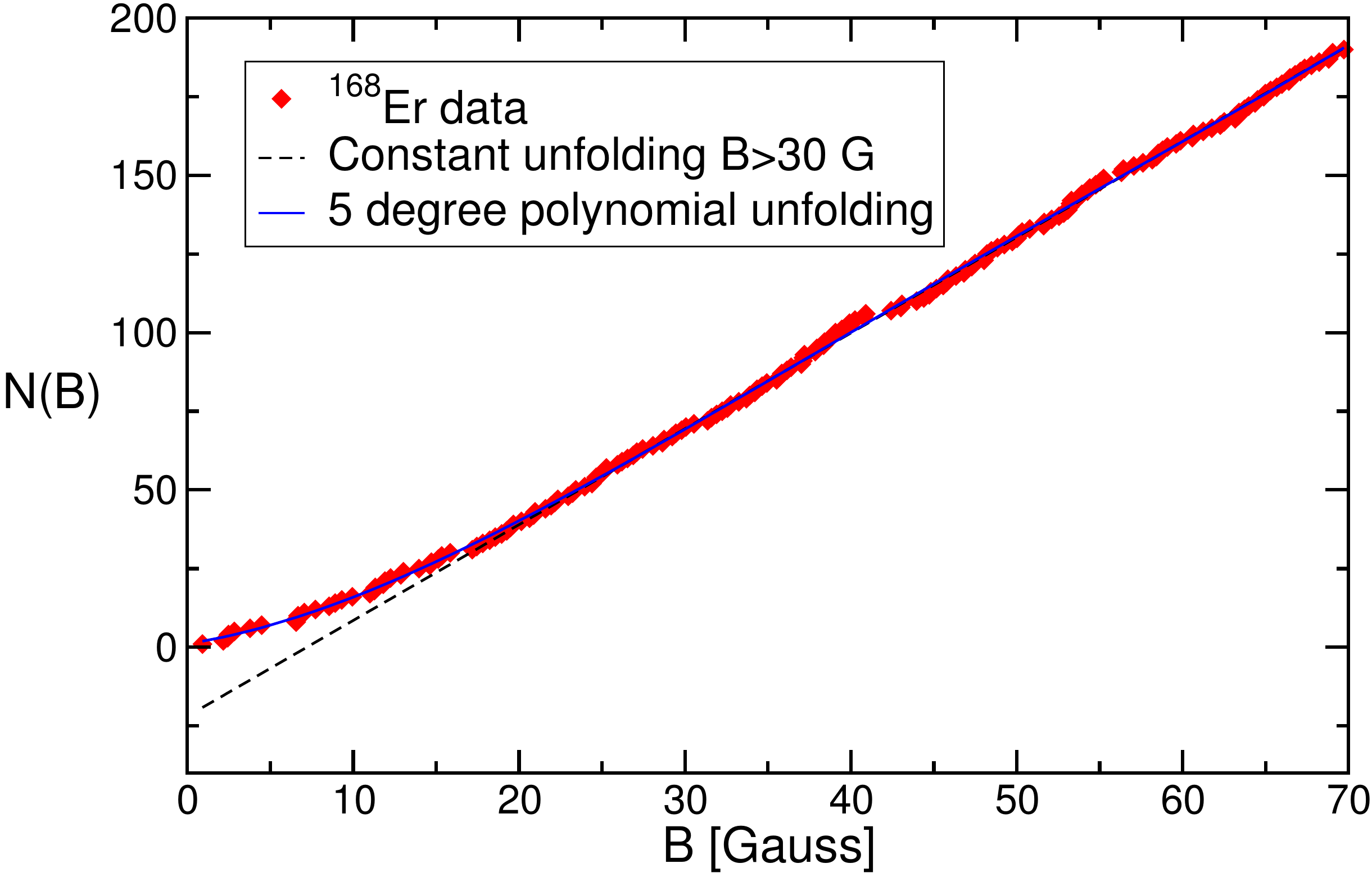}
\caption{\label{fig:unfolding}
 (Color online)
 Comparison of the accumulated number of levels as a function of the magnetic field $N(B)$ for $^{166}$Er. Symbols represent data from Ref.~\cite{Frisch14} while lines are linear (dashed) and polynomial (full) fits. Similar results hold for $^{168}$Er (not shown).}
\end{figure}

We have chosen to use two unfolding methods to cross-check the robustness of our results. The first one is a \textit{local unfolding} where an average density of states is calculated in a window of energies (in the present case, a window of magnetic fields) and this window is moved as we unfold the spectra~\cite{Haake_book}.
This local unfolding gives good results for the short-range correlations such as the nearest-neighbor spacing (NNS) distribution, but it underestimates dramatically long-range correlations \cite{Gomez02}.
The second method is a polynomial fit of the accumulated density of states $N(B)$\textemdash also referred to as the ``staircase function''\textemdash as a function of external magnetic field, which is a measure of the resonance excitation energy.
The simplest case of a constant density of states amounts to a linear fit to the accumulated density of states.
In Fig. \ref{fig:unfolding} we show a comparison of the actual accumulated density of states with the smooth approximation to a fifth order polynomial for all data $0<B<70$G, and a linear fit to data for $B>30$G for both isotopes. The results of the polynomial fit are very good for all magnetic fields.

%%%%%%%%%%%%%%%%%%%%%%%%%%%%%%%%%%%%%%%%%%%%%%%%%%%%%%%%
%%%%%%%%%%%%%%%%%%%%%%%%%%%%%%%%%%%%%%%%%%%%%%%%%%%%%%%%
\subsection{Short range correlations}\label{ssec:shortrange}

Once a reliable unfolding has been made we can start studying spectral statistics. The most used tool for comparing with RMT results is the nearest-neighbor spacing (NNS) distribution, $P(s)$, that measures short range correlations and, in particular, the level repulsion between spectra.
The NNSs are defined as the differences between consecutive unfolded energies $s_i=\epsilon_{i+1}-\epsilon_i$, so that the mean $\langle s\rangle=1$.
We have checked that both local and polynomial unfoldings give essentially the same results for $P(s)$.

In order to study level repulsion we fit the results of the $P(s)$ to the Brody distribution~\cite{Brody73}, as was made in the original experimental paper~\cite{Frisch14}. The Brody distribution is a phenomenological distribution that interpolates as a function of the repulsion parameter $\beta$ between the Poisson distribution corresponding to regular spectra (when $\beta=0$) and the Wigner-Dyson (WD) distribution corresponding to chaotic spectra ($\beta=1$). Its use allows to assess whether the NNS distribution of a series of levels agrees with the prediction of RMT \textit{assuming} that all levels have been observed. It is given by the expression
\begin{align}
P_{\textrm{Brody}}(s,\beta) &=
\alpha(\beta)^{\beta+1}s^{\beta} e^{-\alpha(\beta)s^{\beta+1}} \:,
\label{eq:Brody}
\end{align}
where
$\alpha(\beta) = \Gamma\left(\frac{\beta+2}{\beta+1}\right)$.
Reference~\cite{Frisch14} reported a global study, involving all data points at fields $B>30$~G, which pointed to an intermediate behavior between Poisson and WD.
To analyze the chaoticity of the system as a function of the external magnetic field (or excitation energy), and in particular the possibility that chaos sets in only starting from a certain value of the magnetic field, we have also performed an energy-resolved study: We have divided the spectra in windows of 60 levels around a particular magnetic field and calculated the corresponding Brody parameter.
We note that sixty levels is not a large enough number to retrieve strong statistical evidence from the fit, as we illustrate by means of numerical experiments in the Appendix~\ref{app:fits}.
Nevertheless, calculation of this moving fit to Eq.~\eqref{eq:Brody} provides us with a closer look at the behavior of the system that will permit a more complete understanding of the results that follow.

\begin{figure}[tb]
 \centering
 \includegraphics[width=\linewidth]{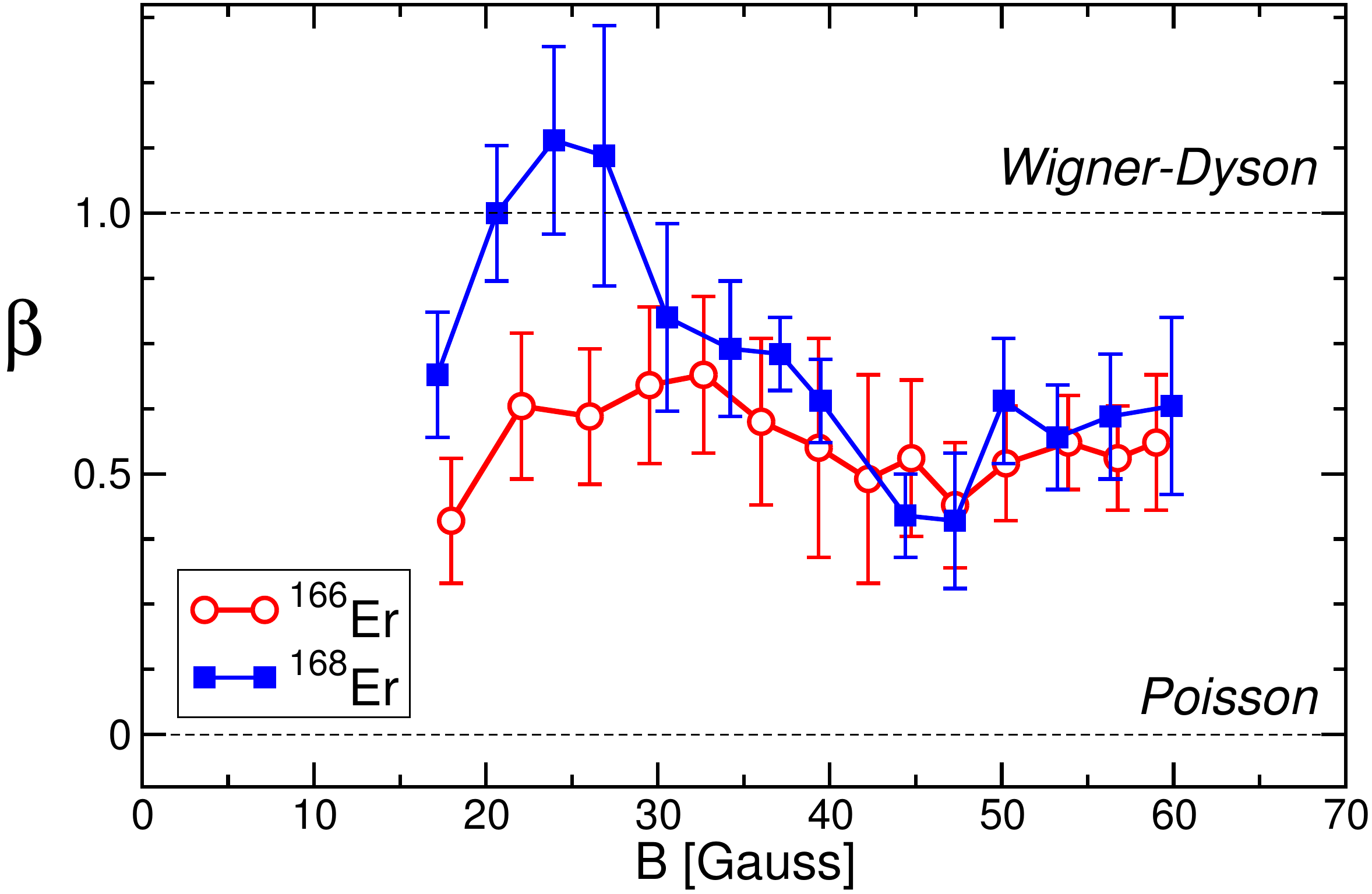}
 \caption{\label{fig:brodyfit}
 (Color online)
 Brody parameter vs. magnetic field from fits of the nearest-neighbor spacing distribution $P(s)$ on a moving window of 60 levels for $^{166}$Er (empty red circles) and $^{168}$Er (filled blue squares). The dashed horizontal lines indicate the limiting values corresponding to Poisson and WD distributions.
 The vertical error bars indicate the errors in the fit.}
\end{figure}

The results of this analysis of short-range fluctuations are shown in Fig.~\ref{fig:brodyfit}.
For the case of $^{166}$Er, we see that $\beta \approx 0.5$ throughout the data set. This seems to point that this isotope is in an intermediate case, not featuring full chaos, although it is difficult to arrive to any conclusions taking into account the errors of the fit. The case of $^{168}$Er seems to be clearer. For magnetic fields $B \lesssim 20$~G the system has not reached full chaos and presents an intermediate behavior between chaos and regularity ($\beta\approx0.7$). Then, for $B \approx 30$ G, we get $\beta\approx1$ and the system appears to reach full chaos. However, for larger values of the magnetic field, the results are hindered by some different problem as $\beta$ falls back to $\approx0.5$. It is hard to imagine any physical mechanism inducing a decrease of chaoticity as the magnetic field is increased. We think that these results are consistent with some small loss of levels for high values of the magnetic field. Indeed, it is known that an incomplete data set from a fully chaotic series will show intermediate features. The analysis of the long-range correlations shown in the next subsection is consistent with this preliminary conclusion and allows us to be more quantitative.

%%%%%%%%%%%%%%%%%%%%%%%%%%%%%%%%%%%%%%%%%%%%%%%%%%%%%%%%
%%%%%%%%%%%%%%%%%%%%%%%%%%%%%%%%%%%%%%%%%%%%%%%%%%%%%%%%
\subsection{Long range correlations}\label{ssec:longrange}

In order to make a complete analysis of the chaoticity of a many-body system, long-range correlations are as important as short-range correlations~\cite{Gomez11}.
There are many different statistics to study long range correlations, they are usually relatively complex to calculate and difficult to analyze statistically. A very useful one because of its simplicity is the power spectrum of the $\delta_q$ statistic. This is defined as
\begin{equation}
\delta_q=\sum_{i=1}^q\left(s_i-\left<s\right>\right) = \epsilon_{q+1}-\epsilon_1-q
\:,
\end{equation}
from which the power spectrum is introduced as~\cite{Relano02,Faleiro06}
\begin{equation}
P_{\delta}(k)=
\frac{1}{L} \left|
  \sum_{q=1}^{L} \delta_q \exp \left( \frac{-2 \pi i k q}{L}\right)
  \right|^2
\label{eq:powerspectrum}
\end{equation}
where $L$ is the total number of levels in the sequence.
As shown in \cite{Relano02,Faleiro04}, 
$P_{\delta}(k) \propto 1/k$ for chaotic systems while $P_{\delta}(k) \propto 1/k^2$ for regular systems, which allows a clear distinction between these two effects.

\begin{figure}[tb]
 \centering
 \includegraphics[width=\linewidth]{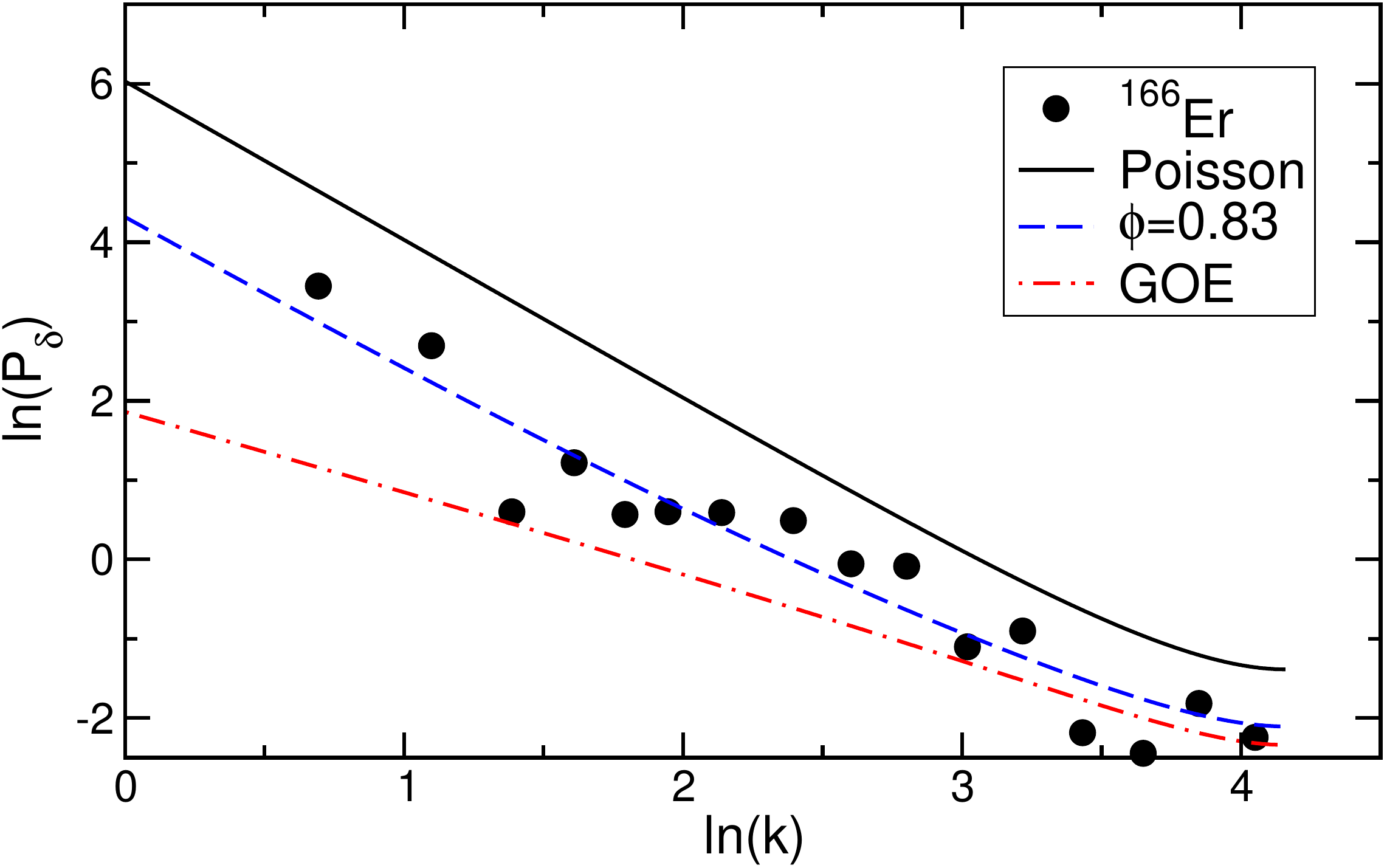} 
 \includegraphics[width=\linewidth]{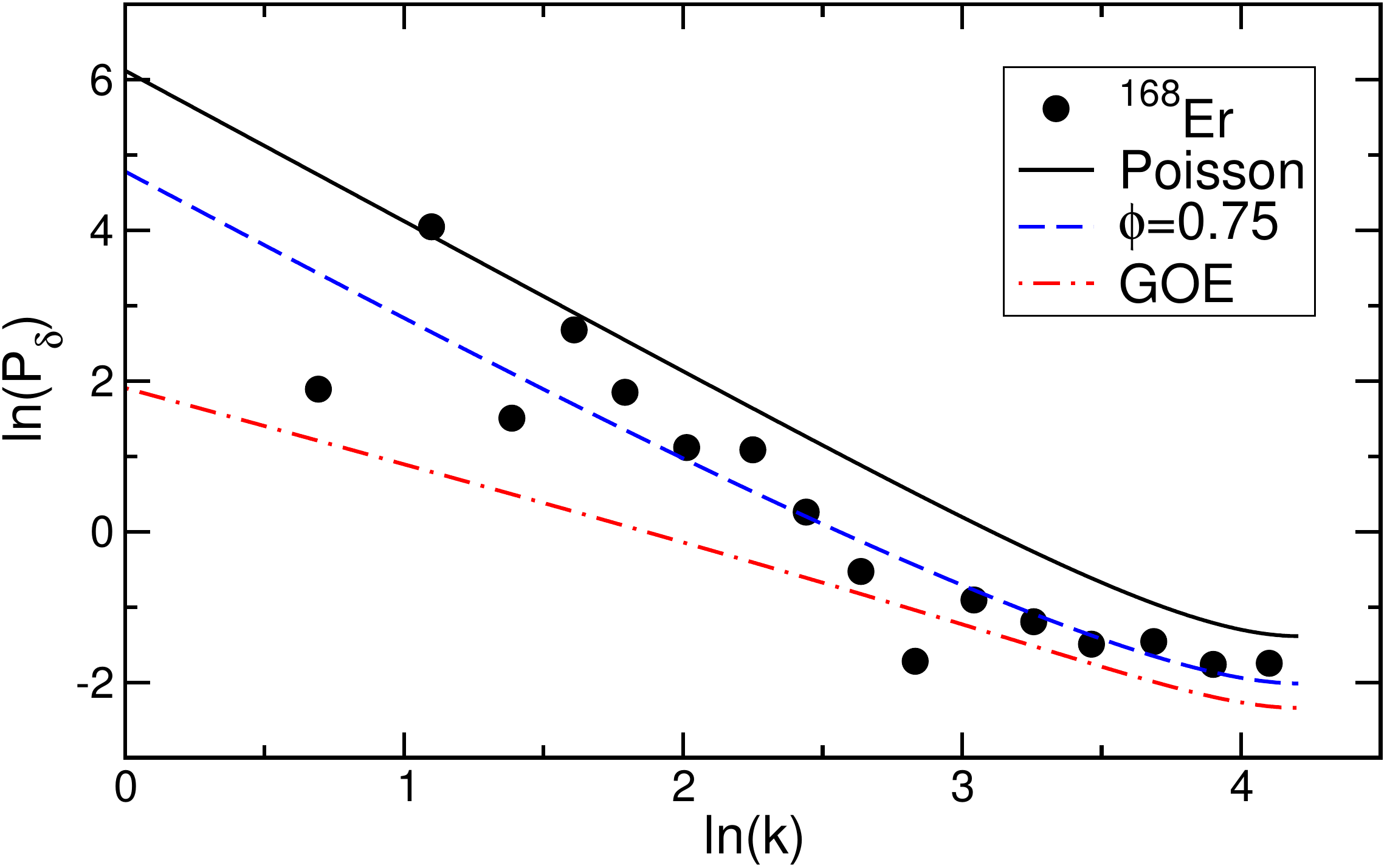}
 \caption{\label{fig:powerspectrum}
 (Color online)
  Power spectrum of the $\delta_q$ for $^{166}$Er (top) and $^{168}$Er (bottom).
 Symbols stand for data from experiments, while lines correspond to the Poisson prediction, a fit to Eq. \eqref{eq:powerspectrum}, and the GOE prediction (from top to bottom in each panel).}
\end{figure}

The results of our numerical studies of the long-range statistics on the experimental data are shown in Fig.~\ref{fig:powerspectrum} as symbols.
In Ref. \cite{Molina07}, a very complete study of the effects of missing levels and mixed symmetries on the long-range correlations of the spectra was made and general formulas for $P_{\delta}(k)$ were derived. Under some simplifying assumptions, these formulas can be adapted to fit relatively small spectra (of the order of 100 levels), as is the case with the present erbium data. 
These assumptions are that the different independent symmetries mixed in the spectra have roughly the same number of levels, that the missing levels are distributed uniformly along the spectra and for the different symmetries, and that the statistical spectral properties of each independent symmetry without any levels missing coincide with the results of the GOE (i.e., each series is fully chaotic). Then, one has a two-parameter formula to fit [see Eq.~(28) in~\cite{Molina07}]. The two parameters are $l$, the number of mixed symmetries, and $\phi$, the observed fraction of levels. As discussed in \cite{Molina07}, as long as the original sequences follow GOE statistics, the values for these parameters as obtained from the fit are good estimations of the real parameters even if the other assumptions of the formula are not strictly fulfilled. 

Figure~\ref{fig:powerspectrum} shows the results of this analysis as lines. It follows that the data for $^{166}$Er at fields $B>30$~G, and for $^{168}$Er at $B>25$~G, are consistent with having only one symmetry present in the spectrum and a fraction of observed levels of $\phi=0.83 \pm 0.08$ for $^{166}$Er and $\phi=0.75 \pm 0.08$ for $^{168}$Er.
The formula for the power spectrum as a function of $\phi$ in the case of only one symmetry present is written in the Appendix, Eq.~\eqref{eq:deltaphi}.
The presence of only one symmetry is expected, as collisions between erbium atoms occur at sufficiently low temperatures ($T\lesssim1$~mK) that only the $s$-wave channel is open. At the same time, due to the substantial anisotropy of the interaction potentials between lanthanide species~\cite{Kotochigova11,Petrov12,FerlainoPC}, collisions couple many possible internal states, with large rotational excitations, that need to be taken into account~\cite{Frisch14}, leading to the complex spectra observed.

In summary, the statistical analysis of the level positions point to $^{168}$Er featuring a chaotic spectrum at fields $B\approx 30$~G, while an apparent $20-25\%$ fraction of missing levels precludes further conclusions at larger fields or for $^{166}$Er.
In the following section, we will perform an independent study of the level widths of both isotopes, and determine whether this enables to estimate the missing fraction.

%%%%%%%%%%%%%%%%%%%%%%%%%%%%%%%%%%%%%%%%%%%%%%%%%%%%%%%%
%%%%%%%%%%%%%%%%%%%%%%%%%%%%%%%%%%%%%%%%%%%%%%%%%%%%%%%%
%%%%%%%%%%%%%%%%%%%%%%%%%%%%%%%%%%%%%%%%%%%%%%%%%%%%%%%%

In principle, it is possible to estimate the fraction of missing levels from the short-range fluctuations as well~\cite{Bohigas06}. Actually, with high-quality data, it should be possible to distinguish between a $P(s)$ intermediate between GOE and Poisson due to missing levels and intermediate statistics with a better fit to a Brody distribution. Yet in practice, as we illustrate in the Appendix, short-range fluctuations are very ill-suited to fit a fraction of missing levels for small level sequences while long-range fluctuations, including the power spectrum, work very well even with sequences as small as $100$ levels.

\section{Statistics of Resonance Widths}\label{sec:widths}

We proceed in this section to perform a statistical analysis of the observed resonance widths.
It is one of the tenets of quantum chaos theory that ``the widths of successive resonances are statistically uncorrelated''~\cite{Watson1981}, in other words, the positions and widths of levels are independent quantities, containing qualitatively different information of the physical system~\cite{Stoeckmann_book}.
For the case of a fully chaotic system, partial widths, $\Delta$, for the decay through $\nu$ channels are expected to be distributed according to a $\chi^2$ distribution with $\nu$ degrees of freedom. This implies that level widths for a series of levels with the same symmetry ($\nu=1$) follow the Porter-Thomas (PT) distribution~\cite{pt1956},
\begin{align}
 \label{eq:pt}
 P_{\textrm{PT}}( \Delta ) &= \sqrt{ \frac{1}{2\pi\Delta\DeltaAver} }
    \exp \left( -\frac{\Delta}{2\DeltaAver} \right) \:,
\end{align}
where $\DeltaAver$ is the average width.
This prediction has been confirmed in a variety of experiments in nuclear~\cite{Shriner1987}, atomic~\cite{Camarda1983}, and molecular spectra~\cite{Zimmermann1988,Delon1991}, as well as with microwave cavities~\cite{Alt1995,Dembowski2005}.

\begin{figure}[tb]
 \centering
 \includegraphics[width=\linewidth]{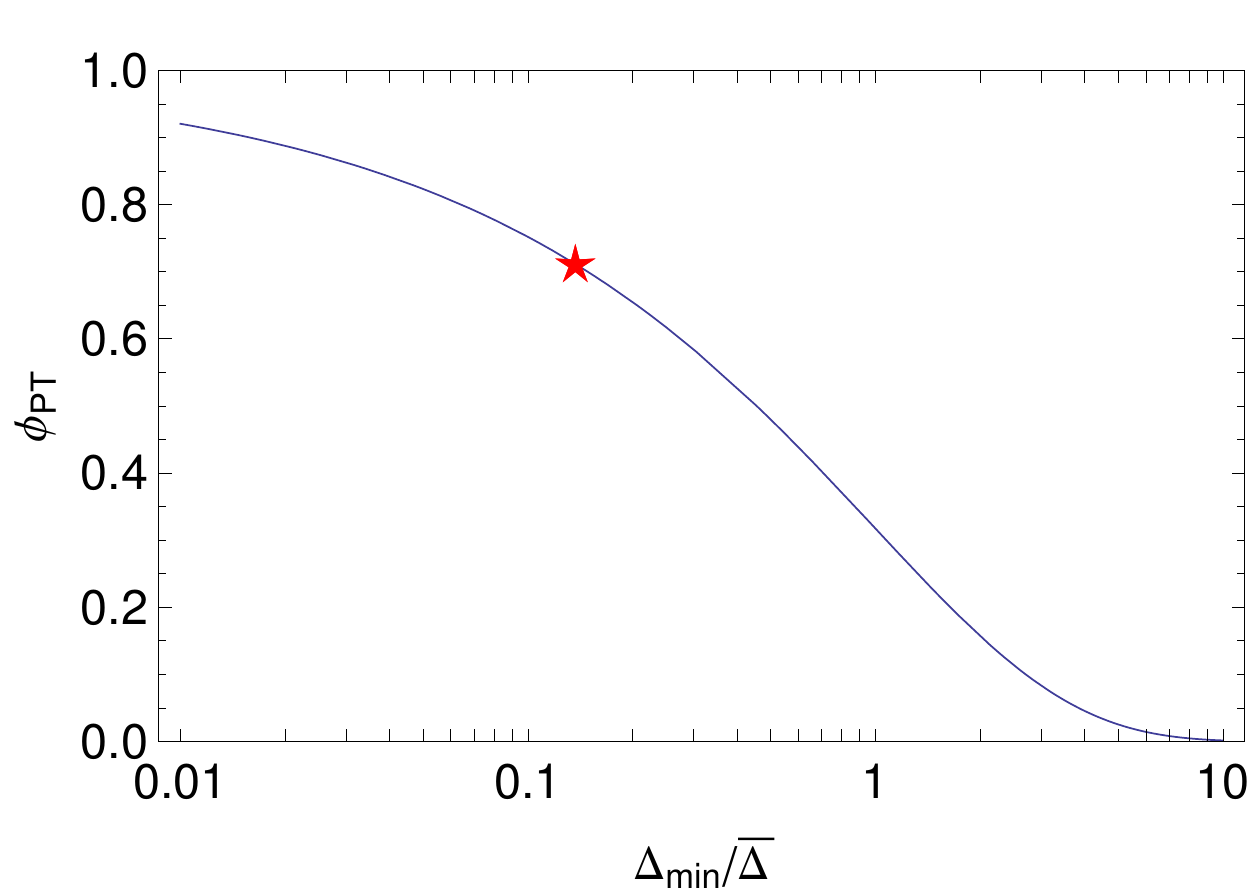}
    % pt_observedfrac.pdf/png calculated in PorterThomas_Flambaum20150622.nb
 \includegraphics[width=\linewidth]{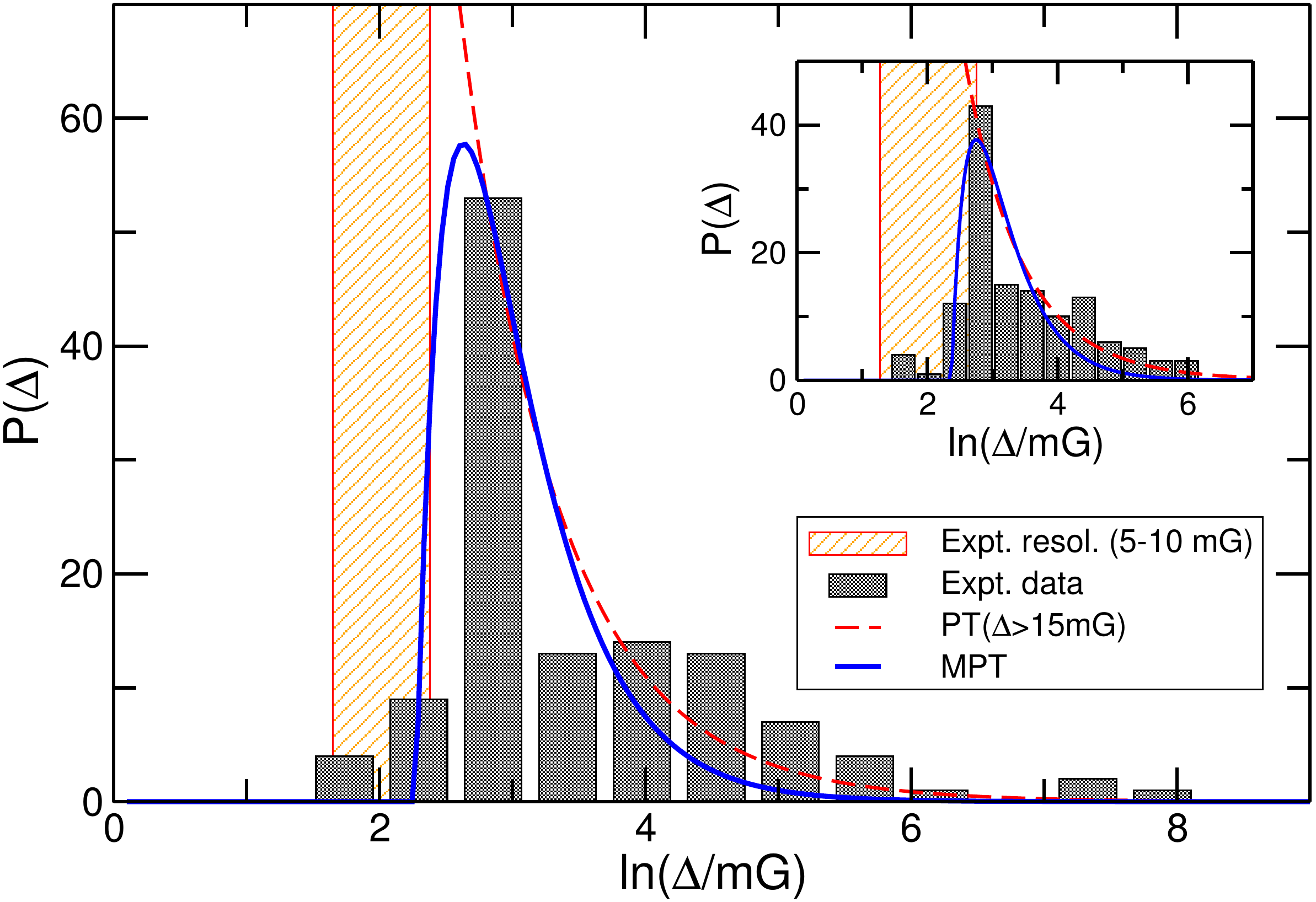}
 \caption{%
  \label{fig:porterthomas}
  (Color online)
  (top) Observed fraction estimated from the PT distribution~\eqref{eq:pt} if only widths $\Delta>\Delta_{\mathrm{min}}$ are observed. The full line is the analytical estimate~\eqref{eq:observedfrac}, the symbol stands for the estimate of the situation corresponding to the experiments in Ref.~\cite{Frisch14} (see text for details).
  (bottom) Probability density of widths, $P(\Delta)$, for $^{168}$Er.
  Probabilities derived from experimental data are shown as shaded bins while
  lines correspond to best fits to the PT~\eqref{eq:pt} (dashed line, $\chidosred=5.634$) and MPT~\eqref{eq:mpt} (solid line, $\chidosred=3.116$) distributions. The hashed rectangle illustrates the range of minimum experimental magnetic field resolution, $\Bresol=5-10$~mG~\cite{Frisch14}.
  The inset shows analogous results for $^{166}$Er with $\chidosred(\mathrm{PT})=6.842$, $\chidosred(\mathrm{MPT})=3.572$.
  }
\end{figure}

Of particular interest for our present study is the possibility to estimate the fraction of missing levels from a comparison of the distribution of level widths with Eq.~\eqref{eq:pt}~\cite{Froehner80,Flambaum98}.
Indeed, if only widths of magnitude $\Delta>\DeltaMin$ are measured, one expects to have an observed fraction given by
\begin{align}
 \label{eq:observedfrac}
 \phi_{\mathrm{PT}} = \int_{\DeltaMin}^{\infty} P_{\mathrm{PT}}(\Delta) d\Delta = \mathrm{erfc}\left( \sqrt{ \frac{\DeltaMin}{2\DeltaAver} } \right) \:.
\end{align}
The dependence of this estimate on $\DeltaMin$ is shown in the upper panel of Fig~\ref{fig:porterthomas}. Here, we have also included an estimate of the observed fraction, using the value of the smallest width reported in~\cite{Frisch14} to estimate $\DeltaMin=4.3$~mG, and the average of observed widths as an estimate of $\DeltaAver=30$~mG~\footnote{To avoid giving too much weight to the few data points with very large widths, we have calculated the average $\overline{\ln(\Delta)}=3.4 \Rightarrow \DeltaAver=\exp(\overline{\ln(\Delta)})=30$~mG.}. From these values, it follows a rough estimate $\phi_{\mathrm{PT}}=71\%$, which is in reasonable agreement with the value obtained in Sec.~\ref{sec:levels} based on level positions.

In the bottom panel of Fig.~\ref{fig:porterthomas} we show fits of the distribution of experimental widths of both Er isotopes to Eq.~\eqref{eq:pt} as dashed lines.
Due to the reduced number of small-width data points, the fit to the PT distribution has been done taking into account only widths larger than 15~mG.
Still, there is a noticeable discrepancy between the best fit to the PT distribution and the experimental data (as goodness-of-fit measures, values for $\chidosred$, the chi-square statistic divided by the number of degrees of freedom, are reported in the caption). Among the various differences, we remark:
\begin{itemize}
\item[(i)] The data do not seem to drop off as quickly as Eq.~\eqref{eq:pt} predicts. As large level widths should be rather scarce, this may be due to statistical fluctuations, that happen to feature ``too many'' broad levels in the relatively small experimental data set. This scenario appears to be rather unlikely as the calculated value of the $\chi^2$ statistic depends mainly on the deviations for large-width levels and the corresponding $p$ values (the probability of obtaining the observed data from a sample following the fitted distribution) are smaller than $0.002$ for all the cases, including the fits to the modified Porter-Thomas distribution explained below.
\item[(ii)] It is possible that some levels with small or intermediate values of $\Delta$ are missing, leading to an underestimation of $P(\Delta)$ away from the maximum observed widths.
\item[(iii)] Compared with analogous results in other quantum-chaotic systems~\cite{Watson1981,Zimmermann1988,Alt1995}, we notice in Fig.~\ref{fig:porterthomas} a striking lack of the divergence signaling that the limiting behavior $P(\Delta) \approx \mathrm{const}/\sqrt{\Delta}$ $(\Delta \ll \DeltaAver)$ has been reached. To recover this feature, it appears that one would need to measure down to much smaller widths than so far observed.
\end{itemize}

To assess which of these effects, if any, is playing the most relevant role, further experimental studies will be needed. At this point, however, we can already note that the average width estimated with the available data, $\DeltaAver\approx30$ mG, is similar to the experimental magnetic field resolution ($\Bresol=5-10$ mG)~\cite{Frisch14}.
Experience from similar studies in other contexts points that one needs a width resolution of order $\approx 10^{-2}\DeltaAver$ to make a faithful comparison with Eq.~\eqref{eq:pt}~\cite{Watson1981,Zimmermann1988,Alt1995}.
Hence, a study of point (iii) would require new experimental measurements with a magnetic field resolution of about $10^{-2}\DeltaAver = 0.3$ mG. Indeed, we note that experiments with another lanthanide species, Dy, already observe an increase in the number of resonant features when the magnetic-field resolution employed is reduced from $5-10$ mG to 1.3 mG~\cite{Baumann14}. Our estimate based on Eq.~\eqref{eq:observedfrac} also points in the same direction.

An alternative method to estimate the values of $\DeltaMin$ and $\DeltaAver$ and, hence, the missing fraction, relies on the modified Porter-Thomas (MPT) distribution introduced by Flambaum \textit{et al.}~\cite{Flambaum98} in their study of electromagnetic $E1$ transitions in Ce atoms. It reads:
\begin{align}
 \label{eq:mpt}
 P_{\textrm{MPT}}( \Delta ) &=
 \left\{ \begin{array}{ll}
   0 & \Delta \leq \DeltaMin \\
   C P_{\mathrm{PT}}(\Delta)
     \left[ 1-\left(\frac{\DeltaMin}{\Delta}\right)^n \right] & \Delta > \DeltaMin
 \end{array}
 \right.     \:,
\end{align}
with $C$ a normalization constant.
Fits of the erbium data to Eq.~\eqref{eq:mpt} are shown as full lines in Fig.~\ref{fig:porterthomas}. They appear to be a much closer representation of the data, especially for widths smaller than the average width, which is reflected in the smaller values of $\chidosred$ than those of fits to the PT distribution.
This puts the focus of missing resonances on those of small width.
However, the highly nonlinear character of the MPT distribution, and the limited statistics available, make it difficult to extract robust information from the fitted parameters. For example, the fit to the $^{168}$Er data yields $\DeltaMin=9.3$~mG, $\DeltaAver_\mathrm{fit}=1.3$~mG.
It is suggestive to note that the fitted value of the parameter $\DeltaMin$ is close to the experimental magnetic field resolution $\Bresol \approx 5-10$~mG. 
An integration similar to Eq.~\eqref{eq:observedfrac} using $P_\mathrm{MPT}$ with the fitted parameters, and setting as the lower limit of integration the largest value of the experimental resolution \Bresol, allows one to estimate the fraction of observed levels with widths $\Delta\geq\max(\Bresol)$:
$\phi_\mathrm{MPT}=\int_{10~\mathrm{mG}}^{\infty} P_{\mathrm{MPT}}(\Delta) d\Delta= 95.5\%$, i.e., less than $5\%$ of levels \textit{with widths $\Delta>\Bresol$} would have not been detected.
On the other hand, as $\DeltaMin>\min(\Bresol)=5$~mG, according to the MPT model, if all data had been collected with the finer magnetic-field resolution, then we would not expect to have missed any level with $\Delta>\Bresol$, and the whole $20-25\%$ missing fraction estimated in Sec.~\ref{ssec:longrange} would originate in narrow resonances, $\Delta<\min(\Bresol)$.

From this evidence, and taking into account our estimation in Sec.~\ref{sec:levels} of a sizable missing fraction, we conclude that most missing levels must have widths $\Delta \lesssim 10$~mG. This implies that a higher-resolution experimental study to address points (ii) and (iii) would be a more direct way to illuminate the role of chaos on erbium molecular resonances, rather than increasing the sample size by exploring higher magnetic fields [point (i)].

Finally, we note the considerable difference between the average width estimated from the experimental data, \DeltaAver, and that coming from the MPT fit, $\DeltaAver_\mathrm{fit}$.
It is worth mentioning that $\DeltaAver$ is only a factor $\approx1/10$ smaller than the mean level spacing ($\approx 310-330$ mG depending on the isotope (cf.~Ref.~\citep{Frisch14}). 
If this were the case, one would expect to have overlapping resonances in the spectrum, making its analysis more complicated.
A recent study of this problem has been reported in~\cite{Fyodorov2015} which shows how the PT distribution needs to be modified in this situation. An analysis of the erbium data following this approach, however, points to a rather weak coupling ($g\approx3.3$) so that differences between the PT distribution and the $P_{M=1}(\Delta)$ distribution in~\cite{Fyodorov2015} are only expected for widths $\gtrsim15$ times larger than the average width.
The erbium data only feature a handful of such points, so that both fits are dominated by resonances with $\Delta\approx \overline{\Delta}$ and yield essentially the same results.
Because of this, one cannot draw any hard conclusions on the existence of such resonance overlaps with the available data.

%%%%%%%%%%%%%%%%%%%%%%%%%%%%%%%%%%%%%%%%%%%%%%%%%%%%%%%%
%%%%%%%%%%%%%%%%%%%%%%%%%%%%%%%%%%%%%%%%%%%%%%%%%%%%%%%%
%%%%%%%%%%%%%%%%%%%%%%%%%%%%%%%%%%%%%%%%%%%%%%%%%%%%%%%%
\section{Conclusions}\label{sec:concl}

In order to study experimentally a quantum system from the point of view of chaos and spectral statistics, a combination of tools is needed to address the questions of chaoticity versus regularity taking into account possible experimental errors both in missing levels and mixing of symmetries. Only results that are consistent throughout different statistics allow to distinguish with confidence between experimental issues and different degrees of chaoticity.
We believe that the analysis with combined tools performed here should become a standard approach for the statistical analysis of experimental spectra. 

We have performed such a comprehensive, multi-faceted analysis for the molecular resonances in $^{166}$Er and $^{168}$Er observed in collision experiments at ultracold temperatures~\cite{Frisch14}. We have analyzed both short range and long range fluctuations in the resonance positions, and the width distribution of the resonances. The conclusion from the spectral fluctuations is that there is a rapid transition to full chaos as a function of the magnetic field for $^{168}$Er, and the data are consistent with a fraction of missing levels of $20-25\%$.
A study of the distribution of experimental widths in comparison with the predictions of RMT including a minimum observable width as a fit parameter, yields an independent estimate of about $5\%$ missing levels above a minimum width about 10~mG; this highlights that the majority of missing levels must have widths below the reported experimental resolution.

In summary, the present study consistently points to a sizable number of narrow resonances having gone undetected in the experiments, and provides specific recommendations for the magnetic field resolution needed to obtain higher-quality data, which are in line with experimental observations of molecular resonances in another rare-earth atomic species~\cite{Baumann14}.

\begin{acknowledgments}

We thank A. Frisch and F. Ferlaino for sharing the experimental data from Ref.~\cite{Frisch14} with us.
This work was supported by Spanish MINECO
projects No.\ FIS2012-33022 and FIS2012-34479,
the JAE-Doc program (CSIC and European Social Fund),
and CAM consortium QUITEMAD+ (SC2013/ICE2801).

\end{acknowledgments}

\appendix

%%%%%%%%%%%%%%%%%%%%%%%%%%%%%%%%%%%%%%%%%%%%%%%%%%%%%%%%
%%%%%%%%%%%%%%%%%%%%%%%%%%%%%%%%%%%%%%%%%%%%%%%%%%%%%%%%
%%%%%%%%%%%%%%%%%%%%%%%%%%%%%%%%%%%%%%%%%%%%%%%%%%%%%%%%
\section{Effect of small sample size on the analysis of spectral statistics}
\label{app:fits}

\begin{figure*}[tb]
 \centering
 \includegraphics[width=\linewidth]{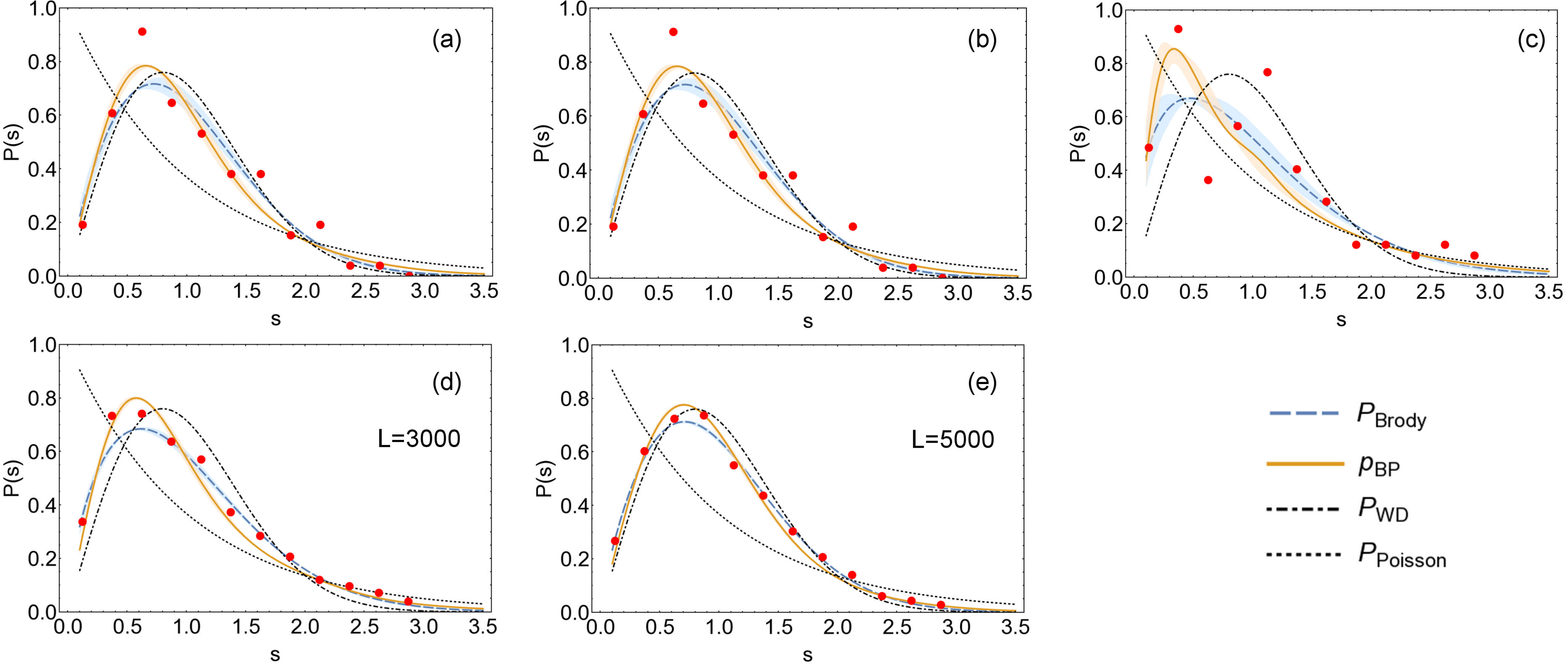}
\caption{\label{fig:samplesize}
 (Color online)
  Short range correlations from three finite samples drawn from a Wigner-Dyson distribution from which $20\%$ of levels have been dropped.
 Symbols stand for the corresponding data, while lines stand for the Poisson prediction, $P(s)=\exp{(-s)}$ (dashedline ), the Wigner surmise (dot-dashed), and fits to Eq.~\eqref{eq:Brody} (long-dashed blue line) and to the model in Ref.~\cite{Bohigas06} (solid orange line). The colored regions show the uncertainty stemming from the fitted values.
 The top panels (a)-(c) correspond to samples with $L=120$ levels.
 The bottom panels corresponds to samples with $L=3000$ (d) and 5000 (e) levels, respectively.
 Fit parameters are:
 (a) $\beta_{\textrm{fit}}=0.90\pm0.15$, $\phi_{\textrm{fit}}=0.95\pm0.10$; % ps8.dat
 (b) $\beta_{\textrm{fit}}=0.81\pm0.11$, $\phi_{\textrm{fit}}=0.78\pm0.07$; % ps6.dat
 (c) $\beta_{\textrm{fit}}=0.43\pm0.16$, $\phi_{\textrm{fit}}=0.66\pm0.10$; % ps4.dat
 (d) $\beta_{\textrm{fit}}=0.62\pm0.05$, $\phi_{\textrm{fit}}=0.34\pm0.04$; % n3000pf05a.dat
 (e) $\beta_{\textrm{fit}}=0.79\pm0.03$, $\phi_{\textrm{fit}}=0.15\pm0.03$. % n5000pf08c.dat
 Note how the uncertainties in the fitted parameters for the small samples (a)-(c) notably underestimate their actual spread.
 }
\end{figure*}

As an illustration of the effect that a reduced sample size ($L\approx 100$ levels) can have on the estimation of the Brody parameter, we show in Figs.~\ref{fig:samplesize}(a-c) numerical results of a fit to Eq.~\eqref{eq:Brody} of three sets of $L=120$ data points constructed by first creating a sample of 150 points drawn randomly with the WD distribution, from which then $20\%$ of points (uniformly distributed within the sample) are discarded.
The data set on the first panel is best fitted with $\beta_{\textrm{fit}}=0.90$, i.e., it appears to be almost fully chaotic.
The central panel estimates $\beta_{\textrm{fit}}=0.81$,
while in the right panel $\beta_{\textrm{fit}}=0.43$.
The figures also contain fits according to the procedure devised in Ref.~\cite{Bohigas06}, which allows to estimate the fraction of missing levels assuming the data correspond to a fully chaotic series. Using this method, we have estimated for each case observed fractions of $\phi_{\textrm{fit}}=95\%$, 78\%, and 34\%, respectively. Note how, especially for the last data series, the two fitting functions show greatly different behaviors\textemdash an effect here purely due to the reduced sample size.

The lower two panels show the results of a similar study with larger samples, $L=3000$ [Fig.~\ref{fig:samplesize}(d)] and 5000 [Fig.~\ref{fig:samplesize}(e)] levels, obtained by discarding $50\%$ and $20\%$ of data points from a similar initial population.
These simulations show that one needs to go to rather large data sets to be able to retrieve with confidence the missing fraction and/or degree of chaoticity from an analysis of these statistical measures.

\begin{figure}[tb]
 \centering
 \includegraphics[width=\linewidth,clip=true]{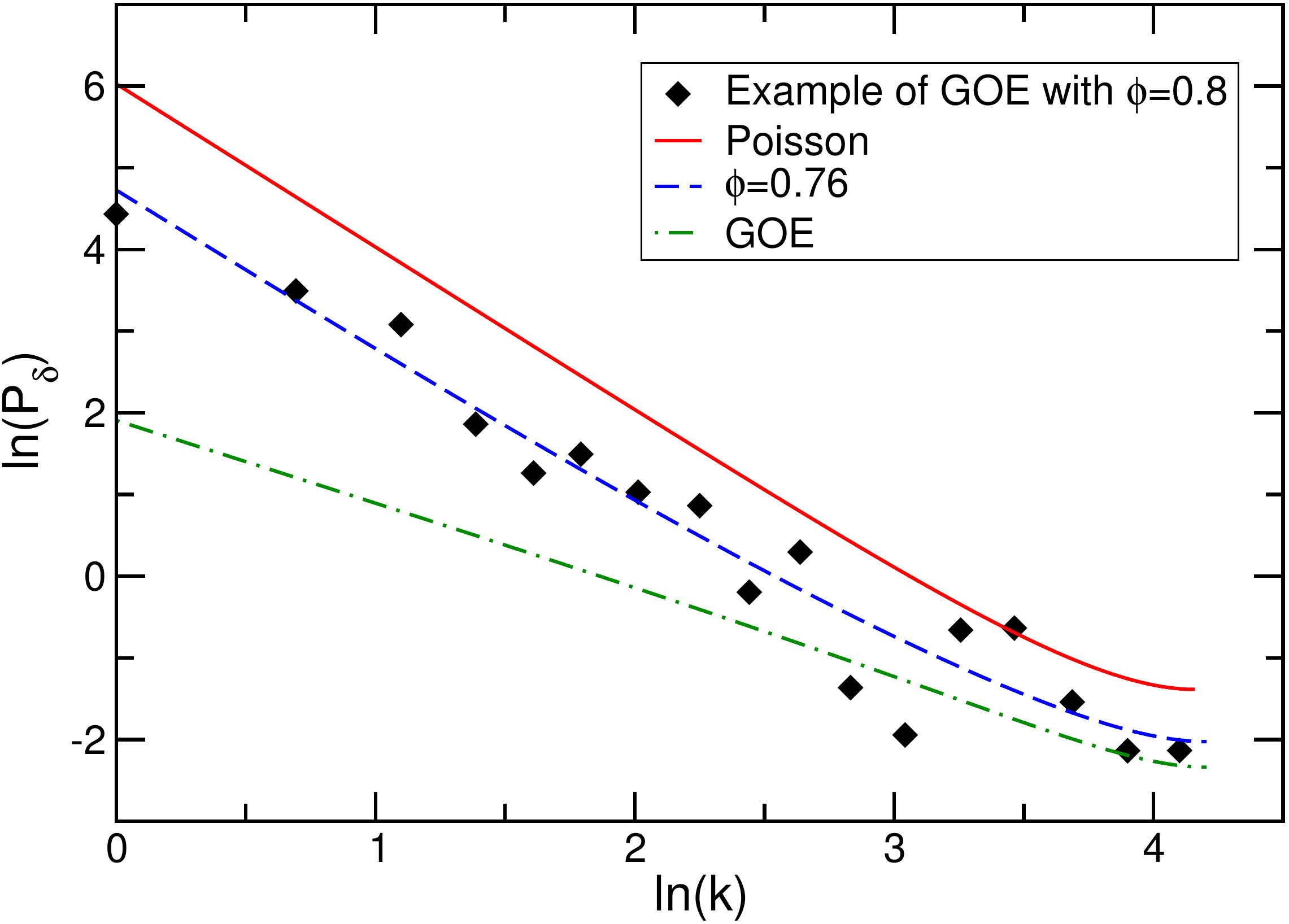}
 \caption{\label{fig:powgoe}
 (Color online)
  Power spectrum for a typical example of a finite sample with $L=135$ levels taken out of a GOE matrix with a fraction of observed levels $\phi=0.8$. The data (symbols) are compared to the fitted result $\phi=0.76$ and the GOE and Poisson predictions.}
\end{figure}

Finally, we performed a similar exercise for long-range correlations and the power spectrum. We calculated a number of examples with $L=135$ levels coming from GOE matrices from which a fraction of $20\%$ of the levels were taken out randomly.
The power spectrum of the $\delta_q$ statistic was calculated and results were fitted to the same formula as we used for fitting the power spectrum of the Er isotopes,
\begin{equation}
 \label{eq:deltaphi}
 P_{\delta}^{\phi}(k)=\frac{\phi n^2}{4 \pi^2}\left[\frac{K\left(\frac{\phi k}{n}\right)-1}{k^2}+\frac{K\left(\frac{\phi (n-k)}{n}\right)-1}{(n-k)^2}\right]-\frac{\phi^2}{12}.
\end{equation}
where $K(t)=2t-t\ln(1+2t)$ is the spectral form factor coming from the GOE, $n=L-1$ is the number of spacings in the sequence, and the parameter $\phi$ is the fraction of observed levels~\cite{Molina07}. The results for the fitting parameter were always within $15\%$ of the actual fraction of
observed levels $\phi=0.8$ and for most cases the real fraction was within the error bars of the fit. A typical example is shown in Fig. \ref{fig:powgoe}.

This numerical exercise shows once more the difficulty to disentangle two situations (experimental missing levels vs.\ system with intermediate chaos), and highlights the need of high-quality data to reach reliable conclusions from finite data series.
A more detailed quantitative analysis of the reliability of different statistics for fitting the fraction of observed levels is outside the scope of this work and will be published elsewhere.

\bibliographystyle{apsrev4-1}
\bibliography{qchaos}

\begin{thebibliography}{41}%
\makeatletter
\providecommand \@ifxundefined [1]{%
 \@ifx{#1\undefined}
}%
\providecommand \@ifnum [1]{%
 \ifnum #1\expandafter \@firstoftwo
 \else \expandafter \@secondoftwo
 \fi
}%
\providecommand \@ifx [1]{%
 \ifx #1\expandafter \@firstoftwo
 \else \expandafter \@secondoftwo
 \fi
}%
\providecommand \natexlab [1]{#1}%
\providecommand \enquote  [1]{``#1''}%
\providecommand \bibnamefont  [1]{#1}%
\providecommand \bibfnamefont [1]{#1}%
\providecommand \citenamefont [1]{#1}%
\providecommand \href@noop [0]{\@secondoftwo}%
\providecommand \href [0]{\begingroup \@sanitize@url \@href}%
\providecommand \@href[1]{\@@startlink{#1}\@@href}%
\providecommand \@@href[1]{\endgroup#1\@@endlink}%
\providecommand \@sanitize@url [0]{\catcode `\\12\catcode `\$12\catcode
  `\&12\catcode `\#12\catcode `\^12\catcode `\_12\catcode `\%12\relax}%
\providecommand \@@startlink[1]{}%
\providecommand \@@endlink[0]{}%
\providecommand \url  [0]{\begingroup\@sanitize@url \@url }%
\providecommand \@url [1]{\endgroup\@href {#1}{\urlprefix }}%
\providecommand \urlprefix  [0]{URL }%
\providecommand \Eprint [0]{\href }%
\providecommand \doibase [0]{http://dx.doi.org/}%
\providecommand \selectlanguage [0]{\@gobble}%
\providecommand \bibinfo  [0]{\@secondoftwo}%
\providecommand \bibfield  [0]{\@secondoftwo}%
\providecommand \translation [1]{[#1]}%
\providecommand \BibitemOpen [0]{}%
\providecommand \bibitemStop [0]{}%
\providecommand \bibitemNoStop [0]{.\EOS\space}%
\providecommand \EOS [0]{\spacefactor3000\relax}%
\providecommand \BibitemShut  [1]{\csname bibitem#1\endcsname}%
\let\auto@bib@innerbib\@empty
%</preamble>
\bibitem [{\citenamefont {Bernath}(2005)}]{Bernath}%
  \BibitemOpen
  \bibfield  {author} {\bibinfo {author} {\bibfnamefont {P.~F.}\ \bibnamefont
  {Bernath}},\ }\href
  {https://global.oup.com/ushe/product/spectra-of-atoms-and-molecules-9780195177596}
  {\emph {\bibinfo {title} {Spectra of Atoms and Molecules}}},\ \bibinfo
  {edition} {2nd}\ ed.\ (\bibinfo  {publisher} {Oxford University Press},\
  \bibinfo {address} {Oxford},\ \bibinfo {year} {2005})\BibitemShut {NoStop}%
\bibitem [{\citenamefont {Brown}\ and\ \citenamefont
  {Carrington}(2003)}]{BrownCarr}%
  \BibitemOpen
  \bibfield  {author} {\bibinfo {author} {\bibfnamefont {J.}~\bibnamefont
  {Brown}}\ and\ \bibinfo {author} {\bibfnamefont {A.}~\bibnamefont
  {Carrington}},\ }\href {\doibase 10.2277/0521530784} {\emph {\bibinfo {title}
  {Rotational Spectroscopy of Diatomic Molecules}}},\ Cambridge Molecular
  Science\ (\bibinfo  {publisher} {Cambridge University Press},\ \bibinfo
  {address} {Cambridge, UK},\ \bibinfo {year} {2003})\BibitemShut {NoStop}%
\bibitem [{\citenamefont {Haake}(2001)}]{Haake_book}%
  \BibitemOpen
  \bibfield  {author} {\bibinfo {author} {\bibfnamefont {F.}~\bibnamefont
  {Haake}},\ }\href {\doibase 10.1007/978-3-662-04506-0} {\emph {\bibinfo
  {title} {Quantum Signatures of Chaos}}},\ \bibinfo {edition} {2nd}\ ed.,\
  \bibinfo {series} {Springer Series in Synergetics}, Vol.~\bibinfo {volume}
  {54}\ (\bibinfo  {publisher} {Sprigner},\ \bibinfo {address} {Berlin},\
  \bibinfo {year} {2001})\BibitemShut {NoStop}%
\bibitem [{\citenamefont {St{\"o}ckmann}(2006)}]{Stoeckmann_book}%
  \BibitemOpen
  \bibfield  {author} {\bibinfo {author} {\bibfnamefont {H.-J.}\ \bibnamefont
  {St{\"o}ckmann}},\ }\href {http://www.cambridge.org/9780521592840} {\emph
  {\bibinfo {title} {Quantum chaos: An introduction}}}\ (\bibinfo  {publisher}
  {Cambridge University Press},\ \bibinfo {address} {Cambridge, UK},\ \bibinfo
  {year} {2006})\BibitemShut {NoStop}%
\bibitem [{\citenamefont {Bohigas}\ \emph {et~al.}(1984)\citenamefont
  {Bohigas}, \citenamefont {Giannoni},\ and\ \citenamefont
  {Schmit}}]{Bohigas84}%
  \BibitemOpen
  \bibfield  {author} {\bibinfo {author} {\bibfnamefont {O.}~\bibnamefont
  {Bohigas}}, \bibinfo {author} {\bibfnamefont {M.~J.}\ \bibnamefont
  {Giannoni}}, \ and\ \bibinfo {author} {\bibfnamefont {C.}~\bibnamefont
  {Schmit}},\ }\href {\doibase 10.1103/PhysRevLett.52.1} {\bibfield  {journal}
  {\bibinfo  {journal} {Phys. Rev. Lett.}\ }\textbf {\bibinfo {volume} {52}},\
  \bibinfo {pages} {1} (\bibinfo {year} {1984})}\BibitemShut {NoStop}%
\bibitem [{\citenamefont {M\"uller}\ \emph {et~al.}(2004)\citenamefont
  {M\"uller}, \citenamefont {Heusler}, \citenamefont {Braun}, \citenamefont
  {Haake},\ and\ \citenamefont {Altland}}]{Mueller04}%
  \BibitemOpen
  \bibfield  {author} {\bibinfo {author} {\bibfnamefont {S.}~\bibnamefont
  {M\"uller}}, \bibinfo {author} {\bibfnamefont {S.}~\bibnamefont {Heusler}},
  \bibinfo {author} {\bibfnamefont {P.}~\bibnamefont {Braun}}, \bibinfo
  {author} {\bibfnamefont {F.}~\bibnamefont {Haake}}, \ and\ \bibinfo {author}
  {\bibfnamefont {A.}~\bibnamefont {Altland}},\ }\href {\doibase
  10.1103/PhysRevLett.93.014103} {\bibfield  {journal} {\bibinfo  {journal}
  {Phys. Rev. Lett.}\ }\textbf {\bibinfo {volume} {93}},\ \bibinfo {pages}
  {014103} (\bibinfo {year} {2004})}\BibitemShut {NoStop}%
\bibitem [{\citenamefont {G\'omez}\ \emph {et~al.}(2011)\citenamefont
  {G\'omez}, \citenamefont {Kar}, \citenamefont {Kota}, \citenamefont {Molina},
  \citenamefont {{Rela\~no}},\ and\ \citenamefont {Retamosa}}]{Gomez11}%
  \BibitemOpen
  \bibfield  {author} {\bibinfo {author} {\bibfnamefont {J.~M.~G.}\
  \bibnamefont {G\'omez}}, \bibinfo {author} {\bibfnamefont {K.}~\bibnamefont
  {Kar}}, \bibinfo {author} {\bibfnamefont {V.~K.~B.}\ \bibnamefont {Kota}},
  \bibinfo {author} {\bibfnamefont {R.~A.}\ \bibnamefont {Molina}}, \bibinfo
  {author} {\bibfnamefont {A.}~\bibnamefont {{Rela\~no}}}, \ and\ \bibinfo
  {author} {\bibfnamefont {J.}~\bibnamefont {Retamosa}},\ }\href {\doibase
  10.1016/j.physrep.2010.11.003} {\bibfield  {journal} {\bibinfo  {journal}
  {Physics Reports}\ }\textbf {\bibinfo {volume} {499}},\ \bibinfo {pages}
  {103} (\bibinfo {year} {2011})}\BibitemShut {NoStop}%
\bibitem [{\citenamefont {Weidenm\"uller}\ and\ \citenamefont
  {Mitchell}(2009)}]{Weiden09}%
  \BibitemOpen
  \bibfield  {author} {\bibinfo {author} {\bibfnamefont {H.~A.}\ \bibnamefont
  {Weidenm\"uller}}\ and\ \bibinfo {author} {\bibfnamefont {G.~E.}\
  \bibnamefont {Mitchell}},\ }\href {\doibase 10.1103/RevModPhys.81.539}
  {\bibfield  {journal} {\bibinfo  {journal} {Rev. Mod. Phys.}\ }\textbf
  {\bibinfo {volume} {81}},\ \bibinfo {pages} {539} (\bibinfo {year}
  {2009})}\BibitemShut {NoStop}%
\bibitem [{\citenamefont {Pique}\ \emph {et~al.}(1987)\citenamefont {Pique},
  \citenamefont {Chen}, \citenamefont {Field},\ and\ \citenamefont
  {Kinsey}}]{Pique1987}%
  \BibitemOpen
  \bibfield  {author} {\bibinfo {author} {\bibfnamefont {J.~P.}\ \bibnamefont
  {Pique}}, \bibinfo {author} {\bibfnamefont {Y.}~\bibnamefont {Chen}},
  \bibinfo {author} {\bibfnamefont {R.~W.}\ \bibnamefont {Field}}, \ and\
  \bibinfo {author} {\bibfnamefont {J.~L.}\ \bibnamefont {Kinsey}},\ }\href
  {\doibase 10.1103/PhysRevLett.58.475} {\bibfield  {journal} {\bibinfo
  {journal} {Phys. Rev. Lett.}\ }\textbf {\bibinfo {volume} {58}},\ \bibinfo
  {pages} {475} (\bibinfo {year} {1987})}\BibitemShut {NoStop}%
\bibitem [{\citenamefont {Zimmermann}\ \emph {et~al.}(1988)\citenamefont
  {Zimmermann}, \citenamefont {K\"oppel}, \citenamefont {Cederbaum},
  \citenamefont {Persch},\ and\ \citenamefont {Demtr\"oder}}]{Zimmermann1988}%
  \BibitemOpen
  \bibfield  {author} {\bibinfo {author} {\bibfnamefont {T.}~\bibnamefont
  {Zimmermann}}, \bibinfo {author} {\bibfnamefont {H.}~\bibnamefont
  {K\"oppel}}, \bibinfo {author} {\bibfnamefont {L.~S.}\ \bibnamefont
  {Cederbaum}}, \bibinfo {author} {\bibfnamefont {G.}~\bibnamefont {Persch}}, \
  and\ \bibinfo {author} {\bibfnamefont {W.}~\bibnamefont {Demtr\"oder}},\
  }\href {\doibase 10.1103/PhysRevLett.61.3} {\bibfield  {journal} {\bibinfo
  {journal} {Phys. Rev. Lett.}\ }\textbf {\bibinfo {volume} {61}},\ \bibinfo
  {pages} {3} (\bibinfo {year} {1988})}\BibitemShut {NoStop}%
\bibitem [{\citenamefont {Weiner}\ \emph {et~al.}(1999)\citenamefont {Weiner},
  \citenamefont {Bagnato}, \citenamefont {Zilio},\ and\ \citenamefont
  {Julienne}}]{weiner1999}%
  \BibitemOpen
  \bibfield  {author} {\bibinfo {author} {\bibfnamefont {J.}~\bibnamefont
  {Weiner}}, \bibinfo {author} {\bibfnamefont {V.~S.}\ \bibnamefont {Bagnato}},
  \bibinfo {author} {\bibfnamefont {S.}~\bibnamefont {Zilio}}, \ and\ \bibinfo
  {author} {\bibfnamefont {P.~S.}\ \bibnamefont {Julienne}},\ }\href {\doibase
  10.1103/RevModPhys.71.1} {\bibfield  {journal} {\bibinfo  {journal} {Rev.
  Mod. Phys.}\ }\textbf {\bibinfo {volume} {71}},\ \bibinfo {pages} {1}
  (\bibinfo {year} {1999})}\BibitemShut {NoStop}%
\bibitem [{\citenamefont {Stuhl}\ \emph {et~al.}(2014)\citenamefont {Stuhl},
  \citenamefont {Hummon},\ and\ \citenamefont {Ye}}]{stuhl2014}%
  \BibitemOpen
  \bibfield  {author} {\bibinfo {author} {\bibfnamefont {B.~K.}\ \bibnamefont
  {Stuhl}}, \bibinfo {author} {\bibfnamefont {M.~T.}\ \bibnamefont {Hummon}}, \
  and\ \bibinfo {author} {\bibfnamefont {J.}~\bibnamefont {Ye}},\ }\href
  {\doibase 10.1146/annurev-physchem-040513-103744} {\bibfield  {journal}
  {\bibinfo  {journal} {Annual review of physical chemistry}\ }\textbf
  {\bibinfo {volume} {65}},\ \bibinfo {pages} {501} (\bibinfo {year}
  {2014})}\BibitemShut {NoStop}%
\bibitem [{\citenamefont {Krems}(2008)}]{krems2008}%
  \BibitemOpen
  \bibfield  {author} {\bibinfo {author} {\bibfnamefont {R.~V.}\ \bibnamefont
  {Krems}},\ }\href {\doibase 10.1039/b802322k} {\bibfield  {journal} {\bibinfo
   {journal} {Phys. Chem. Chem. Phys.}\ }\textbf {\bibinfo {volume} {10}},\
  \bibinfo {pages} {4079} (\bibinfo {year} {2008})}\BibitemShut {NoStop}%
\bibitem [{\citenamefont {Qu\'em\'ener}\ and\ \citenamefont
  {Julienne}(2012)}]{quemener2012}%
  \BibitemOpen
  \bibfield  {author} {\bibinfo {author} {\bibfnamefont {G.}~\bibnamefont
  {Qu\'em\'ener}}\ and\ \bibinfo {author} {\bibfnamefont {P.~S.}\ \bibnamefont
  {Julienne}},\ }\href {\doibase 10.1021/cr300092g} {\bibfield  {journal}
  {\bibinfo  {journal} {Chemical Reviews}\ }\textbf {\bibinfo {volume} {112}},\
  \bibinfo {pages} {4949} (\bibinfo {year} {2012})}\BibitemShut {NoStop}%
\bibitem [{\citenamefont {Mur-Petit}\ \emph {et~al.}(2012)\citenamefont
  {Mur-Petit}, \citenamefont {Garc\'{i}a-Ripoll}, \citenamefont
  {P\'erez-R\'{i}os}, \citenamefont {Campos-Mart\'{i}nez}, \citenamefont
  {Hern\'andez},\ and\ \citenamefont {Willitsch}}]{murpetit2012}%
  \BibitemOpen
  \bibfield  {author} {\bibinfo {author} {\bibfnamefont {J.}~\bibnamefont
  {Mur-Petit}}, \bibinfo {author} {\bibfnamefont {J.~J.}\ \bibnamefont
  {Garc\'{i}a-Ripoll}}, \bibinfo {author} {\bibfnamefont {J.}~\bibnamefont
  {P\'erez-R\'{i}os}}, \bibinfo {author} {\bibfnamefont {J.}~\bibnamefont
  {Campos-Mart\'{i}nez}}, \bibinfo {author} {\bibfnamefont {M.~I.}\
  \bibnamefont {Hern\'andez}}, \ and\ \bibinfo {author} {\bibfnamefont
  {S.}~\bibnamefont {Willitsch}},\ }\href {\doibase 10.1103/PhysRevA.85.022308}
  {\bibfield  {journal} {\bibinfo  {journal} {Phys. Rev. A}\ }\textbf {\bibinfo
  {volume} {85}},\ \bibinfo {pages} {022308} (\bibinfo {year}
  {2012})}\BibitemShut {NoStop}%
\bibitem [{\citenamefont {Mur-Petit}\ and\ \citenamefont
  {Garc\'{i}a-Ripoll}(2015)}]{murpetit2015}%
  \BibitemOpen
  \bibfield  {author} {\bibinfo {author} {\bibfnamefont {J.}~\bibnamefont
  {Mur-Petit}}\ and\ \bibinfo {author} {\bibfnamefont {J.~J.}\ \bibnamefont
  {Garc\'{i}a-Ripoll}},\ }\href {\doibase 10.1103/PhysRevA.91.012504}
  {\bibfield  {journal} {\bibinfo  {journal} {Phys. Rev. A}\ }\textbf {\bibinfo
  {volume} {91}},\ \bibinfo {pages} {012504} (\bibinfo {year}
  {2015})}\BibitemShut {NoStop}%
\bibitem [{\citenamefont {Frisch}\ \emph {et~al.}(2014)\citenamefont {Frisch},
  \citenamefont {Mark}, \citenamefont {Aikawa}, \citenamefont {Ferlaino},
  \citenamefont {Bohn}, \citenamefont {Makrides}, \citenamefont {Petrov},\ and\
  \citenamefont {Kotochigova}}]{Frisch14}%
  \BibitemOpen
  \bibfield  {author} {\bibinfo {author} {\bibfnamefont {A.}~\bibnamefont
  {Frisch}}, \bibinfo {author} {\bibfnamefont {M.}~\bibnamefont {Mark}},
  \bibinfo {author} {\bibfnamefont {K.}~\bibnamefont {Aikawa}}, \bibinfo
  {author} {\bibfnamefont {F.}~\bibnamefont {Ferlaino}}, \bibinfo {author}
  {\bibfnamefont {J.}~\bibnamefont {Bohn}}, \bibinfo {author} {\bibfnamefont
  {C.}~\bibnamefont {Makrides}}, \bibinfo {author} {\bibfnamefont
  {A.}~\bibnamefont {Petrov}}, \ and\ \bibinfo {author} {\bibfnamefont
  {S.}~\bibnamefont {Kotochigova}},\ }\href {\doibase 10.1038/nature1313714}
  {\bibfield  {journal} {\bibinfo  {journal} {Nature}\ }\textbf {\bibinfo
  {volume} {507}},\ \bibinfo {pages} {475 } (\bibinfo {year}
  {2014})}\BibitemShut {NoStop}%
\bibitem [{\citenamefont {Liou}\ \emph {et~al.}(1972)\citenamefont {Liou},
  \citenamefont {Camarda},\ and\ \citenamefont {Rahn}}]{Liou72}%
  \BibitemOpen
  \bibfield  {author} {\bibinfo {author} {\bibfnamefont {H.~I.}\ \bibnamefont
  {Liou}}, \bibinfo {author} {\bibfnamefont {H.~S.}\ \bibnamefont {Camarda}}, \
  and\ \bibinfo {author} {\bibfnamefont {F.}~\bibnamefont {Rahn}},\ }\href
  {\doibase 10.1103/PhysRevC.5.1002} {\bibfield  {journal} {\bibinfo  {journal}
  {Phys. Rev. C}\ }\textbf {\bibinfo {volume} {5}},\ \bibinfo {pages} {1002}
  (\bibinfo {year} {1972})}\BibitemShut {NoStop}%
\bibitem [{\citenamefont {Bohigas}\ and\ \citenamefont
  {Pato}(2006)}]{Bohigas06}%
  \BibitemOpen
  \bibfield  {author} {\bibinfo {author} {\bibfnamefont {O.}~\bibnamefont
  {Bohigas}}\ and\ \bibinfo {author} {\bibfnamefont {M.~P.}\ \bibnamefont
  {Pato}},\ }\href {\doibase 10.1103/PhysRevE.74.036212} {\bibfield  {journal}
  {\bibinfo  {journal} {Phys. Rev. E}\ }\textbf {\bibinfo {volume} {74}},\
  \bibinfo {pages} {036212} (\bibinfo {year} {2006})}\BibitemShut {NoStop}%
\bibitem [{\citenamefont {Molina}\ \emph {et~al.}(2007)\citenamefont {Molina},
  \citenamefont {Retamosa}, \citenamefont {{Mu\~noz}}, \citenamefont
  {{Rela\~no}},\ and\ \citenamefont {Faleiro}}]{Molina07}%
  \BibitemOpen
  \bibfield  {author} {\bibinfo {author} {\bibfnamefont {R.~A.}\ \bibnamefont
  {Molina}}, \bibinfo {author} {\bibfnamefont {J.}~\bibnamefont {Retamosa}},
  \bibinfo {author} {\bibfnamefont {L.}~\bibnamefont {{Mu\~noz}}}, \bibinfo
  {author} {\bibfnamefont {A.}~\bibnamefont {{Rela\~no}}}, \ and\ \bibinfo
  {author} {\bibfnamefont {E.}~\bibnamefont {Faleiro}},\ }\href {\doibase
  10.1016/j.physletb.2006.10.058} {\bibfield  {journal} {\bibinfo  {journal}
  {Physics Letters B}\ }\textbf {\bibinfo {volume} {644}},\ \bibinfo {pages}
  {25 } (\bibinfo {year} {2007})}\BibitemShut {NoStop}%
\bibitem [{\citenamefont {Molina}\ \emph {et~al.}(2000)\citenamefont {Molina},
  \citenamefont {G\'omez},\ and\ \citenamefont {Retamosa}}]{Molina00}%
  \BibitemOpen
  \bibfield  {author} {\bibinfo {author} {\bibfnamefont {R.~A.}\ \bibnamefont
  {Molina}}, \bibinfo {author} {\bibfnamefont {J.~M.~G.}\ \bibnamefont
  {G\'omez}}, \ and\ \bibinfo {author} {\bibfnamefont {J.}~\bibnamefont
  {Retamosa}},\ }\href {\doibase 10.1103/PhysRevC.63.014311} {\bibfield
  {journal} {\bibinfo  {journal} {Phys. Rev. C}\ }\textbf {\bibinfo {volume}
  {63}},\ \bibinfo {pages} {014311} (\bibinfo {year} {2000})}\BibitemShut
  {NoStop}%
\bibitem [{\citenamefont {Fr\"ohner}(1980)}]{Froehner80}%
  \BibitemOpen
  \bibfield  {author} {\bibinfo {author} {\bibfnamefont {F.}~\bibnamefont
  {Fr\"ohner}},\ }\href
  {https://www-nds.iaea.org/publications/tecdocs/iaea-smr-0043/} {\emph
  {\bibinfo {title} {Nuclear Theory for Applications}}},\ \bibinfo {type}
  {Tech. Rep.}\ \bibinfo {number} {IAEA-SMR-43}\ (\bibinfo  {institution}
  {IAEA},\ \bibinfo {address} {Vienna},\ \bibinfo {year} {1980})\ \bibinfo
  {note} {pp. 59-95}\BibitemShut {NoStop}%
\bibitem [{\citenamefont {G\'omez}\ \emph {et~al.}(2002)\citenamefont
  {G\'omez}, \citenamefont {Molina}, \citenamefont {{Rela\~no}},\ and\
  \citenamefont {Retamosa}}]{Gomez02}%
  \BibitemOpen
  \bibfield  {author} {\bibinfo {author} {\bibfnamefont {J.~M.~G.}\
  \bibnamefont {G\'omez}}, \bibinfo {author} {\bibfnamefont {R.~A.}\
  \bibnamefont {Molina}}, \bibinfo {author} {\bibfnamefont {A.}~\bibnamefont
  {{Rela\~no}}}, \ and\ \bibinfo {author} {\bibfnamefont {J.}~\bibnamefont
  {Retamosa}},\ }\href {\doibase 10.1103/PhysRevE.66.036209} {\bibfield
  {journal} {\bibinfo  {journal} {Phys. Rev. E}\ }\textbf {\bibinfo {volume}
  {66}},\ \bibinfo {pages} {036209} (\bibinfo {year} {2002})}\BibitemShut
  {NoStop}%
\bibitem [{\citenamefont {Brody}(1973)}]{Brody73}%
  \BibitemOpen
  \bibfield  {author} {\bibinfo {author} {\bibfnamefont {T.~A.}\ \bibnamefont
  {Brody}},\ }\href {\doibase 10.1007/BF02727859} {\bibfield  {journal}
  {\bibinfo  {journal} {Lett. Nuovo Cimento}\ }\textbf {\bibinfo {volume}
  {7}},\ \bibinfo {pages} {482} (\bibinfo {year} {1973})}\BibitemShut {NoStop}%
\bibitem [{\citenamefont {{Rela\~no}}\ \emph {et~al.}(2002)\citenamefont
  {{Rela\~no}}, \citenamefont {G\'omez}, \citenamefont {Molina}, \citenamefont
  {Retamosa},\ and\ \citenamefont {Faleiro}}]{Relano02}%
  \BibitemOpen
  \bibfield  {author} {\bibinfo {author} {\bibfnamefont {A.}~\bibnamefont
  {{Rela\~no}}}, \bibinfo {author} {\bibfnamefont {J.~M.~G.}\ \bibnamefont
  {G\'omez}}, \bibinfo {author} {\bibfnamefont {R.~A.}\ \bibnamefont {Molina}},
  \bibinfo {author} {\bibfnamefont {J.}~\bibnamefont {Retamosa}}, \ and\
  \bibinfo {author} {\bibfnamefont {E.}~\bibnamefont {Faleiro}},\ }\href
  {\doibase 10.1103/PhysRevLett.89.244102} {\bibfield  {journal} {\bibinfo
  {journal} {Phys. Rev. Lett.}\ }\textbf {\bibinfo {volume} {89}},\ \bibinfo
  {pages} {244102} (\bibinfo {year} {2002})}\BibitemShut {NoStop}%
\bibitem [{\citenamefont {Faleiro}\ \emph {et~al.}(2006)\citenamefont
  {Faleiro}, \citenamefont {Kuhl}, \citenamefont {Molina}, \citenamefont
  {{Mu\~noz}}, \citenamefont {{Rela\~no}},\ and\ \citenamefont
  {Retamosa}}]{Faleiro06}%
  \BibitemOpen
  \bibfield  {author} {\bibinfo {author} {\bibfnamefont {E.}~\bibnamefont
  {Faleiro}}, \bibinfo {author} {\bibfnamefont {U.}~\bibnamefont {Kuhl}},
  \bibinfo {author} {\bibfnamefont {R.~A.}\ \bibnamefont {Molina}}, \bibinfo
  {author} {\bibfnamefont {L.}~\bibnamefont {{Mu\~noz}}}, \bibinfo {author}
  {\bibfnamefont {A.}~\bibnamefont {{Rela\~no}}}, \ and\ \bibinfo {author}
  {\bibfnamefont {J.}~\bibnamefont {Retamosa}},\ }\href {\doibase
  http://dx.doi.org/10.1016/j.physleta.2006.05.029} {\bibfield  {journal}
  {\bibinfo  {journal} {Physics Letters A}\ }\textbf {\bibinfo {volume}
  {358}},\ \bibinfo {pages} {251 } (\bibinfo {year} {2006})}\BibitemShut
  {NoStop}%
\bibitem [{\citenamefont {Faleiro}\ \emph {et~al.}(2004)\citenamefont
  {Faleiro}, \citenamefont {G\'omez}, \citenamefont {Molina}, \citenamefont
  {{Mu\~noz}}, \citenamefont {{Rela\~no}},\ and\ \citenamefont
  {Retamosa}}]{Faleiro04}%
  \BibitemOpen
  \bibfield  {author} {\bibinfo {author} {\bibfnamefont {E.}~\bibnamefont
  {Faleiro}}, \bibinfo {author} {\bibfnamefont {J.~M.~G.}\ \bibnamefont
  {G\'omez}}, \bibinfo {author} {\bibfnamefont {R.~A.}\ \bibnamefont {Molina}},
  \bibinfo {author} {\bibfnamefont {L.}~\bibnamefont {{Mu\~noz}}}, \bibinfo
  {author} {\bibfnamefont {A.}~\bibnamefont {{Rela\~no}}}, \ and\ \bibinfo
  {author} {\bibfnamefont {J.}~\bibnamefont {Retamosa}},\ }\href {\doibase
  10.1103/PhysRevLett.93.244101} {\bibfield  {journal} {\bibinfo  {journal}
  {Phys. Rev. Lett.}\ }\textbf {\bibinfo {volume} {93}},\ \bibinfo {pages}
  {244101} (\bibinfo {year} {2004})}\BibitemShut {NoStop}%
\bibitem [{\citenamefont {Kotochigova}\ and\ \citenamefont
  {Petrov}(2011)}]{Kotochigova11}%
  \BibitemOpen
  \bibfield  {author} {\bibinfo {author} {\bibfnamefont {S.}~\bibnamefont
  {Kotochigova}}\ and\ \bibinfo {author} {\bibfnamefont {A.}~\bibnamefont
  {Petrov}},\ }\href {\doibase 10.1039/C1CP21175G} {\bibfield  {journal}
  {\bibinfo  {journal} {Phys. Chem. Chem. Phys.}\ }\textbf {\bibinfo {volume}
  {13}},\ \bibinfo {pages} {19165} (\bibinfo {year} {2011})}\BibitemShut
  {NoStop}%
\bibitem [{\citenamefont {Petrov}\ \emph {et~al.}(2012)\citenamefont {Petrov},
  \citenamefont {Tiesinga},\ and\ \citenamefont {Kotochigova}}]{Petrov12}%
  \BibitemOpen
  \bibfield  {author} {\bibinfo {author} {\bibfnamefont {A.}~\bibnamefont
  {Petrov}}, \bibinfo {author} {\bibfnamefont {E.}~\bibnamefont {Tiesinga}}, \
  and\ \bibinfo {author} {\bibfnamefont {S.}~\bibnamefont {Kotochigova}},\
  }\href {\doibase 10.1103/PhysRevLett.109.103002} {\bibfield  {journal}
  {\bibinfo  {journal} {Phys. Rev. Lett.}\ }\textbf {\bibinfo {volume} {109}},\
  \bibinfo {pages} {103002} (\bibinfo {year} {2012})}\BibitemShut {NoStop}%
\bibitem [{\citenamefont {Ferlaino}(2015)}]{FerlainoPC}%
  \BibitemOpen
  \bibfield  {author} {\bibinfo {author} {\bibfnamefont {F.}~\bibnamefont
  {Ferlaino}},\ }\href@noop {} {}\bibinfo {howpublished} {private
  communication} (\bibinfo {year} {2015})\BibitemShut {NoStop}%
\bibitem [{\citenamefont {Watson}\ \emph {et~al.}(1981)\citenamefont {Watson},
  \citenamefont {Bilpuch},\ and\ \citenamefont {Mitchell}}]{Watson1981}%
  \BibitemOpen
  \bibfield  {author} {\bibinfo {author} {\bibfnamefont {W.~A.}\ \bibnamefont
  {Watson}}, \bibinfo {author} {\bibfnamefont {E.~G.}\ \bibnamefont {Bilpuch}},
  \ and\ \bibinfo {author} {\bibfnamefont {G.~E.}\ \bibnamefont {Mitchell}},\
  }\href {\doibase 10.1016/0029-554X(81)90268-8} {\bibfield  {journal}
  {\bibinfo  {journal} {Nuclear Instruments and Methods in Physics Research}\
  }\textbf {\bibinfo {volume} {188}},\ \bibinfo {pages} {571} (\bibinfo {year}
  {1981})}\BibitemShut {NoStop}%
\bibitem [{\citenamefont {Porter}\ and\ \citenamefont {Thomas}(1956)}]{pt1956}%
  \BibitemOpen
  \bibfield  {author} {\bibinfo {author} {\bibfnamefont {C.~E.}\ \bibnamefont
  {Porter}}\ and\ \bibinfo {author} {\bibfnamefont {R.~G.}\ \bibnamefont
  {Thomas}},\ }\href {\doibase 10.1103/PhysRev.104.483} {\bibfield  {journal}
  {\bibinfo  {journal} {Phys. Rev.}\ }\textbf {\bibinfo {volume} {104}},\
  \bibinfo {pages} {483} (\bibinfo {year} {1956})}\BibitemShut {NoStop}%
\bibitem [{\citenamefont {Shriner}\ \emph {et~al.}(1987)\citenamefont
  {Shriner}, \citenamefont {Mitchell},\ and\ \citenamefont
  {Bilpuch}}]{Shriner1987}%
  \BibitemOpen
  \bibfield  {author} {\bibinfo {author} {\bibfnamefont {J.~F.}\ \bibnamefont
  {Shriner}}, \bibinfo {author} {\bibfnamefont {G.~E.}\ \bibnamefont
  {Mitchell}}, \ and\ \bibinfo {author} {\bibfnamefont {E.~G.}\ \bibnamefont
  {Bilpuch}},\ }\href {\doibase 10.1103/PhysRevLett.59.435} {\bibfield
  {journal} {\bibinfo  {journal} {Phys. Rev. Lett.}\ }\textbf {\bibinfo
  {volume} {59}},\ \bibinfo {pages} {435} (\bibinfo {year} {1987})}\BibitemShut
  {NoStop}%
\bibitem [{\citenamefont {Camarda}\ and\ \citenamefont
  {Georgopulos}(1983)}]{Camarda1983}%
  \BibitemOpen
  \bibfield  {author} {\bibinfo {author} {\bibfnamefont {H.~S.}\ \bibnamefont
  {Camarda}}\ and\ \bibinfo {author} {\bibfnamefont {P.~D.}\ \bibnamefont
  {Georgopulos}},\ }\href {\doibase 10.1103/PhysRevLett.50.492} {\bibfield
  {journal} {\bibinfo  {journal} {Phys. Rev. Lett.}\ }\textbf {\bibinfo
  {volume} {50}},\ \bibinfo {pages} {492} (\bibinfo {year} {1983})}\BibitemShut
  {NoStop}%
\bibitem [{\citenamefont {Delon}\ \emph {et~al.}(1991)\citenamefont {Delon},
  \citenamefont {Jost},\ and\ \citenamefont {Lombardi}}]{Delon1991}%
  \BibitemOpen
  \bibfield  {author} {\bibinfo {author} {\bibfnamefont {A.}~\bibnamefont
  {Delon}}, \bibinfo {author} {\bibfnamefont {R.}~\bibnamefont {Jost}}, \ and\
  \bibinfo {author} {\bibfnamefont {M.}~\bibnamefont {Lombardi}},\ }\href
  {\doibase 10.1063/1.461620} {\bibfield  {journal} {\bibinfo  {journal} {The
  Journal of Chemical Physics}\ }\textbf {\bibinfo {volume} {95}},\ \bibinfo
  {pages} {5701} (\bibinfo {year} {1991})}\BibitemShut {NoStop}%
\bibitem [{\citenamefont {Alt}\ \emph {et~al.}(1995)\citenamefont {Alt},
  \citenamefont {Gr\"af}, \citenamefont {Harney}, \citenamefont {Hofferbert},
  \citenamefont {Lengeler}, \citenamefont {Richter}, \citenamefont {Schardt},\
  and\ \citenamefont {Weidenm\"uller}}]{Alt1995}%
  \BibitemOpen
  \bibfield  {author} {\bibinfo {author} {\bibfnamefont {H.}~\bibnamefont
  {Alt}}, \bibinfo {author} {\bibfnamefont {H.-D.}\ \bibnamefont {Gr\"af}},
  \bibinfo {author} {\bibfnamefont {H.~L.}\ \bibnamefont {Harney}}, \bibinfo
  {author} {\bibfnamefont {R.}~\bibnamefont {Hofferbert}}, \bibinfo {author}
  {\bibfnamefont {H.}~\bibnamefont {Lengeler}}, \bibinfo {author}
  {\bibfnamefont {A.}~\bibnamefont {Richter}}, \bibinfo {author} {\bibfnamefont
  {P.}~\bibnamefont {Schardt}}, \ and\ \bibinfo {author} {\bibfnamefont
  {H.~A.}\ \bibnamefont {Weidenm\"uller}},\ }\href {\doibase
  10.1103/PhysRevLett.74.62} {\bibfield  {journal} {\bibinfo  {journal} {Phys.
  Rev. Lett.}\ }\textbf {\bibinfo {volume} {74}},\ \bibinfo {pages} {62}
  (\bibinfo {year} {1995})}\BibitemShut {NoStop}%
\bibitem [{\citenamefont {Dembowski}\ \emph {et~al.}(2005)\citenamefont
  {Dembowski}, \citenamefont {Dietz}, \citenamefont {Friedrich}, \citenamefont
  {Gr\"af}, \citenamefont {Harney}, \citenamefont {Heine}, \citenamefont
  {Miski-Oglu},\ and\ \citenamefont {Richter}}]{Dembowski2005}%
  \BibitemOpen
  \bibfield  {author} {\bibinfo {author} {\bibfnamefont {C.}~\bibnamefont
  {Dembowski}}, \bibinfo {author} {\bibfnamefont {B.}~\bibnamefont {Dietz}},
  \bibinfo {author} {\bibfnamefont {T.}~\bibnamefont {Friedrich}}, \bibinfo
  {author} {\bibfnamefont {H.-D.}\ \bibnamefont {Gr\"af}}, \bibinfo {author}
  {\bibfnamefont {H.~L.}\ \bibnamefont {Harney}}, \bibinfo {author}
  {\bibfnamefont {A.}~\bibnamefont {Heine}}, \bibinfo {author} {\bibfnamefont
  {M.}~\bibnamefont {Miski-Oglu}}, \ and\ \bibinfo {author} {\bibfnamefont
  {A.}~\bibnamefont {Richter}},\ }\href {\doibase 10.1103/PhysRevE.71.046202}
  {\bibfield  {journal} {\bibinfo  {journal} {Phys. Rev. E}\ }\textbf {\bibinfo
  {volume} {71}},\ \bibinfo {pages} {046202} (\bibinfo {year}
  {2005})}\BibitemShut {NoStop}%
\bibitem [{\citenamefont {Flambaum}\ \emph {et~al.}(1998)\citenamefont
  {Flambaum}, \citenamefont {Gribakina},\ and\ \citenamefont
  {Gribakin}}]{Flambaum98}%
  \BibitemOpen
  \bibfield  {author} {\bibinfo {author} {\bibfnamefont {V.~V.}\ \bibnamefont
  {Flambaum}}, \bibinfo {author} {\bibfnamefont {A.~A.}\ \bibnamefont
  {Gribakina}}, \ and\ \bibinfo {author} {\bibfnamefont {G.~F.}\ \bibnamefont
  {Gribakin}},\ }\href {\doibase 10.1103/PhysRevA.58.230} {\bibfield  {journal}
  {\bibinfo  {journal} {Phys. Rev. A}\ }\textbf {\bibinfo {volume} {58}},\
  \bibinfo {pages} {230} (\bibinfo {year} {1998})}\BibitemShut {NoStop}%
\bibitem [{Note1()}]{Note1}%
  \BibitemOpen
  \bibinfo {note} {To avoid giving too much weight to the few data points with
  very large widths, we have calculated the average $\protect \overline
  {\protect \qopname \relax o{ln}(\Delta )}=3.4 \Rightarrow \protect
  \ensuremath {\protect \overline {\Delta }}=\protect \qopname \relax
  o{exp}(\protect \overline {\protect \qopname \relax o{ln}(\Delta
  )})=30$~mG.}\BibitemShut {Stop}%
\bibitem [{\citenamefont {Baumann}\ \emph {et~al.}(2014)\citenamefont
  {Baumann}, \citenamefont {Burdick}, \citenamefont {Lu},\ and\ \citenamefont
  {Lev}}]{Baumann14}%
  \BibitemOpen
  \bibfield  {author} {\bibinfo {author} {\bibfnamefont {K.}~\bibnamefont
  {Baumann}}, \bibinfo {author} {\bibfnamefont {N.~Q.}\ \bibnamefont
  {Burdick}}, \bibinfo {author} {\bibfnamefont {M.}~\bibnamefont {Lu}}, \ and\
  \bibinfo {author} {\bibfnamefont {B.~L.}\ \bibnamefont {Lev}},\ }\href
  {\doibase 10.1103/PhysRevA.89.020701} {\bibfield  {journal} {\bibinfo
  {journal} {Phys. Rev. A}\ }\textbf {\bibinfo {volume} {89}},\ \bibinfo
  {pages} {020701} (\bibinfo {year} {2014})}\BibitemShut {NoStop}%
\bibitem [{\citenamefont {Fyodorov}\ and\ \citenamefont
  {Savin}(2015)}]{Fyodorov2015}%
  \BibitemOpen
  \bibfield  {author} {\bibinfo {author} {\bibfnamefont {Y.~V.}\ \bibnamefont
  {Fyodorov}}\ and\ \bibinfo {author} {\bibfnamefont {D.~V.}\ \bibnamefont
  {Savin}},\ }\href {\doibase 10.1209/0295-5075/110/40006} {\bibfield
  {journal} {\bibinfo  {journal} {EPL (Europhysics Letters)}\ }\textbf
  {\bibinfo {volume} {110}},\ \bibinfo {pages} {40006} (\bibinfo {year}
  {2015})}\BibitemShut {NoStop}%
\end{thebibliography}%

%%%%%%%%%%%%%%%%%%%%%%%%%%%%
%%%%%%%%%%%%%%%%%%%%%%%%%%%%
%%%%%%%%%%%%%%%%%%%%%%%%%%%%
%%%%%%%%%%%%%%%%%%%%%%%%%%%%
\end{document}